\documentclass[12pt]{article}

\usepackage{amsmath,amsfonts,amssymb}

\begin{document}

\begin{titlepage}

\begin{center}

\begin{center}
{\Large{ \bf Canonical Description of  T-duality for Fundamental
String and D1-Brane and Double Wick Rotation}}
\end{center}

\vskip 1cm

{\large Josef Kluso\v{n}$^{}$\footnote{E-mail: {\tt
klu@physics.muni.cz}} }

\vskip 0.8cm

{\it Department of
Theoretical Physics and Astrophysics\\
Faculty of Science, Masaryk University\\
Kotl\'{a}\v{r}sk\'{a} 2, 611 37, Brno\\
Czech Republic\\
[10mm]}

\vskip 0.8cm

\end{center}

\begin{abstract}
We study T-duality transformations in canonical formalism for
Nambu-Gotto action. Then we investigate the relation between
world-sheet double Wick rotation and sequence of target space
T-dualities and Wick rotation in case of fundamental string and
D1-brane.

\end{abstract}

\bigskip

\end{titlepage}

\newpage

\newcommand{\mT}{\mathcal{T}}
\def\tr{\mathrm{Tr}}
\def\tb{\tilde{b}}
\def\tJ{\tilde{J}}
\def\str{\mathrm{Str}}
\newcommand{\tL}{\tilde{L}}
\def\tpsi{\tilde{\psi}}
\def\tp{\tilde{p}}
\def\tB{\tilde{B}}
\def\bV{\mathbf{V}}
\def\Pf{\mathrm{Pf}}
\def\ttheta{\tilde{\theta}}
\def\bX{\mathbf{X}}
\def\bY{\mathbf{Y}}
\def\I{\mathbf{i}}
\def\tx{\tilde{x}}
\def\IT{\I_{\Phi,\Phi',T}}
\def\tPi{\tilde{\Pi}}
\def\tmH{\tilde{\mH}}
\def\bC{\mathbf{C}}
\def \cit{\IT^{\dag}}
\def\hPi{\hat{\Pi}}
\def \cdt{\overline{\tilde{D}T}}
\def \dt{\tilde{D}T}
\def\bra #1{\left<#1\right|}
\def\ket #1{\left|#1\right>}
\def\vac #1{\left<\left<#1\right>\right>}
\def\pb  #1{\left\{#1\right\}}
\def \uw #1{(w^{#1})}
\def\bX{\mathbf{X}}
\def\bB{\mathbf{B}}
\def\bC{\mathbf{C}}
\def \dw #1{(w_{#1})}
\newcommand{\bK}{\mathbf{K}}
\newcommand{\thw}{\tilde{\hat{w}}}
\newcommand{\bA}{{\bf A}}
\newcommand{\bd}{{\bf d}}
\newcommand{\bD}{{\bf D}}
\newcommand{\bF}{{\bf F}}
\newcommand{\bN}{{\bf N}}
\newcommand{\hp}{\hat{p}}
\newcommand{\hq}{\hat{q}}
\newcommand{\hF}{\hat{F}}
\newcommand{\hG}{\hat{G}}
\newcommand{\hH}{\hat{H}}
\newcommand{\hU}{\hat{U}}
\newcommand{\mH}{\mathcal{H}}
\newcommand{\mG}{\mathcal{G}}
\newcommand{\mA}{\mathcal{A}}
\newcommand{\mD}{\mathcal{D}}
\newcommand{\tpr}{t^{\prime}}
\newcommand{\bzg}{\overline{\zg}}
\newcommand{\of}{\overline{f}}
\newcommand{\ow}{\overline{w}}
\newcommand{\htheta}{\hat{\theta}}
\newcommand{\opartial}{\overline{\partial}}
\newcommand{\hd}{\hat{d}}
\def\mF{\mathcal{F}}
\newcommand{\halpha}{\hat{\alpha}}
\newcommand{\hbeta}{\hat{\beta}}
\newcommand{\hdelta}{\hat{\delta}}
\newcommand{\hgamma}{\hat{\gamma}}
\newcommand{\hlambda}{\hat{\lambda}}
\newcommand{\hw}{\hat{w}}
\newcommand{\hN}{\hat{N}}
\newcommand{\onabla}{\overline{\nabla}}
\newcommand{\hmu}{\hat{\mu}}
\newcommand{\hnu}{\hat{\nu}}
\newcommand{\ha}{\hat{a}}
\newcommand{\hb}{\hat{b}}
\newcommand{\hc}{\hat{c}}
\newcommand{\com}[1]{\left[#1\right]}
\newcommand{\oz}{\overline{z}}
\newcommand{\oJ}{\overline{J}}
\newcommand{\mL}{\mathcal{L}}
\newcommand{\oh}{\overline{h}}
\newcommand{\oT}{\overline{T}}
\newcommand{\oepsilon}{\overline{\epsilon}}
\newcommand{\tP}{\tilde{P}}
\newcommand{\hP}{\hat{P}}
\newcommand{\talpha}{\tilde{\alpha}}
\newcommand{\uc}{\underline{c}}
\newcommand{\ud}{\underline{d}}
\newcommand{\ue}{\underline{e}}
\newcommand{\uf}{\underline{f}}
\newcommand{\hpi}{\hat{\pi}}
\newcommand{\oZ}{\overline{Z}}
\newcommand{\tg}{\tilde{g}}
\newcommand{\tK}{\tilde{K}}
\newcommand{\tj}{\tilde{j}}
\newcommand{\tG}{\tilde{G}}
\newcommand{\hg}{\hat{g}}
\newcommand{\htG}{\hat{\tilde{G}}}
\newcommand{\hX}{\hat{X}}
\newcommand{\hY}{\hat{Y}}
\newcommand{\bDi}{\left(\bD^{-1}\right)}
\newcommand{\hthteta}{\hat{\theta}}
\newcommand{\hB}{\hat{B}}
\newcommand{\tlambda}{\tilde{\lambda}}
\newcommand{\thlambda}{\tilde{\hat{\lambda}}}
\newcommand{\tw}{\tilde{w}}
\newcommand{\hJ}{\hat{J}}
\newcommand{\tPsi}{\tilde{\Psi}}
\newcommand{\cP}{{\cal P}}
\def\ba{\mathbf{a}}
\newcommand{\tphi}{\tilde{\phi}}
\newcommand{\tOmega}{\tilde{\Omega}}
\newcommand{\homega}{\hat{\omega}}
\newcommand{\hupsilon}{\hat{\upsilon}}
\newcommand{\hUpsilon}{\hat{\Upsilon}}
\newcommand{\hOmega}{\hat{\Omega}}
\newcommand{\bJ}{\mathbf{J}}
\newcommand{\olambda}{\overline{\lambda}}
\newcommand{\uhlambda}{\underline{\hlambda}}
\newcommand{\uhw}{\underline{\hw}}
\newcommand{\tC}{\tilde{C}}
\def \lhw #1{(\hw^{#1})}
\def \dhw #1{(\hw_{#1})}
\newcommand{\bG}{\mathbf{G}}
\newcommand{\bhG}{\hat{\bG}}
\newcommand{\bH}{\mathbf{H}}
\newcommand{\bE}{\mathbf{E}}
\newcommand{\mJ}{\mathcal{J}}
\newcommand{\mY}{\mathcal{Y}}
\newcommand{\mZ}{\mathcal{Z}}
\newcommand{\hj}{\hat{j}}
\newcommand{\bAi}{\left(\bA^{-1}\right)}
\newcommand{\hpartial}{\hat{\partial}}
\newcommand{\hD}{\hat{D}}
\newcommand{\mC}{\mathcal{C}}
\newcommand{\omC}{\overline{\mC}}
\newcommand{\mP}{\mathcal{P}}
\newcommand{\omP}{\overline{\mP}}
\newcommand{\tLambda}{\tilde{\Lambda}}
\newcommand{\tPhi}{\tilde{\Phi}}
\newcommand{\tD}{\tilde{D}}
\newcommand{\tgamma}{\tilde{\gamma}}
%%%%%%%%%%%%%%%%%%%%%
%%%%Introduction %%%%%%%%%
%%%%%%%%%%%%%%%%%%%%
\section{Introduction and Summary }
A double Wick rotation on the string world-sheet was proposed in
\cite{Arutyunov:2007tc} and now it is well known that it is a very
useful tool for the study of AdS/CFT correspondence, for review see
\cite{Arutyunov:2009ga}. It is crucial that double Wick rotation has
sense in case of the light cone gauge fixed string since only in
this case the double Wick rotation transforms string world-sheet
theory non-trivially. In fact, the thermodynamics of this different
two dimensional quantum field theory gives the solution of the
spectral problem of AdS/CFT when the spectrum of the string on
$AdS_5\times S^5$ is computed with the help of the thermodynamic
Bethe ansatz for this mirror model
\cite{Arutyunov:2009zu,Arutyunov:2009ur,Bombardelli:2009ns,Gromov:2009bc}.

Due to  the significance of the double Wick rotation one can ask the
question  whether this transformation has more physical meaning.
Such a question was firstly posed
 in \cite{Arutyunov:2014cra} and it was shown there  that the double Wick
 rotation on the world-sheet of the gauge fixed bosonic string is
 equivalent to the particular transformation of the metric and NS-NS
 two form. This analysis was then extended to the case of the uniform gauge fixed
 Green-Schwarz superstring in \cite{Arutyunov:2014jfa} where it was
 shown that the double Wick rotation is equivalent to the particular
 transformation of metric, NS-NS two form and
  dilaton and Ramond-Ramond fields. It was also
 shown there that these transformations can be interpreted from a
 target space perspective as a combination of $T$ dualities and
 analytic continuition, following A. Tsytlin's suggestion.

Due to the fact that the  idea that the double Wick rotation could
have deeper physical meaning is very attractive  we applied it for
another two dimensional theory which is a low energy effective
action for D1-brane in \cite{Kluson:2015saa}. In this paper we
firstly determined   uniform gauge fixed action for D1-brane in
general background. Then we applied double Wick rotation for this
theory and we found that the double Wick rotated action has the same
form as the original one when the background gravitational and NS-NS
two form fields transform as in \cite{Arutyunov:2014cra} while we
found that dilaton and Ramond-Ramond fields transform differently.
We suggested that this discrepancy could be explained by the fact
that D1-brane behaves differently under T-duality transformations
but we leaved the detailed analysis of this issue for future. The
aim of this paper is to answer this question. More precisely, we
would like to see how suggested combinations of T-dualities and Wick
rotation in the target space-time can be performed on the
world-sheet of the fundamental string and D1-brane action. Due to
the fact that the gauge fixing is important for the  given procedure
and since this gauge fixing is performed when we fix one of the
components of momenta rather than target space coordinate we perform
T-duality transformation on the level of the Hamiltonian formalism.
In fact, the interpretation of T-duality transformation as canonical
transformation was given in \cite{Alvarez:1994wj} \footnote{For
review, see \cite{Alvarez:1994dn}.} in case of gauge fixed string
action. We generalize this analysis to the case of the Nambu-Gotto
(NG) action which is invariant under two dimensional diffeomorphism
and we obtain celebrated Buscher's  rules for the T-duality
transformations of  the background metric and NS-NS two form field
\cite{Buscher:1987sk,Buscher:1987qj} \footnote{We would like to
stress that given procedure could be equivalently performed with
Polyakov form of the string action. Explicitly, the Hamiltonian has
the same form as in case of NG string with subtle difference that
now the primary constraints are the momenta conjugate to the
components of the world-sheet metric. Then the requirement of their
preservations gives the secondary constraints $\mH_\tau,\mH_\sigma$.
Note that these constraints are the primary constraints of the NG
string.}.

 Using this result we perform
sequence of T-duality transformations and Wick rotation in the
target space-time and we show that when we impose the uniform gauge
in the resulting Hamiltonian we obtain that the Hamiltonian for the
physical degrees of freedom is the same as in case of the double
Wick rotated gauge fixed theory with an important exception that the
sequence of T-duality transformation and Wick rotation do not
generate a change $B_{\mu\nu}\rightarrow -B_{\mu\nu}$ where
$B_{\mu\nu}$ are  components of NS-NS two form fields that are
transverse to the directions where corresponding T-duality and Wick
rotations were performed. On the other hand it is well known that
the transformation $B\rightarrow -B$ is the symmetry of supergravity
equations of motion with the absence of RR fields. So that whenever
string propagates on background with $g,B$ that solves the
supergravity equations of motion it can propagates on the background
with $g,-B$. So that without lost of consistency we can augment the
sequence T-duality and Wick rotation with an additional operation
$B\rightarrow -B$. Then this extended sequence of operation  is
equivalent to the double Wick rotation of the uniform gauge fixed
string theory\footnote{The discussion is more subtle in case of
non-zero RR fields and for detailed discussion see
\cite{Arutyunov:2014jfa}.}.

As the next step we extend this analysis to the case of  D1-brane
effective action which consists of Dirac-Born-Infeld (DBI) action
and Wess-Zummino (WZ) term.  We firstly apply the canonical
transformations for this  theory and we derive how the background
fields transform. However we should stress that we are able to
perform such an analysis when we presume that the electric flux is
fixed. This is a natural requirement since it is well known that the
electric flux is proportional to the number of the fundamental
strings and we would like to compare two actions when the number of
fundamental strings is the same. We find that under this canonical
transformation the dilation and Ramond-Ramond fields transform
differently than we should expect from T-duality rules which has
very simple explanation. We argue that the final action corresponds
to the fundamental string action moving in T-dual background when
however the components of the background fields that appear in
Buscher's rules correspond to the S-dual background when we use the
equivalence of the D1-brane action (with constant electric flux) and
fundamental string action in  S-dual background
\cite{Tseytlin:1996it}.

 Due to this
fact we can now explain the discrepancy that we found in the
previous paper \cite{Kluson:2015saa}. Explicitly, we perform the
sequence of canonical transformations and Wick rotation in case of
Hamiltonian for D1-brane. Then we perform the uniform gauge fixing
and we again find that the Hamiltonian for the physical degrees of
freedom implies the transformation rules for the target space fields
that have the same form as in case of the double Wick rotation
performed on the uniform gauge fixed D1-brane effective action again
with an exception that components of Ramond-Ramond two forms that
are transverse to the directions of duality transformations  do not
transform. On the other hand we can again argue for the existence of
the symmetry $C^{(2)}\rightarrow -C^{(2)}, B\rightarrow -B$ of the
solutions of the supergravity equations of motion so that we find
exact equivalence between double Wick rotation of the uniform gauge
fixed D1-brane action and sequence of "canonical
transformation-target space double Wick rotation-canonical
transformation-$(C^{2}\rightarrow -C^{(2)} \ , B\rightarrow -B)$".

%It is important to stress that even if the canonical transformation
%used in the case of D1-brane effective theory has the same form as
%the canonical transformation used in case of the fundamental string
%their physical interpretation is different. In fact, this result
%shows the limitation of the effective field theory description of
%D1-brane in case of T-duality transformations since it is well known
%that the fundamental definition of D-brane is as the hypersurface
%where the open string ends \cite{Polchinski:1995mt}.
% Of course, it is well known how to deal with Dp-brane
%under dimensional reduction in the background with one isometry
%direction. In this case we presume that Dp-brane is wrapping given
%direction so that we can impose the condition that all world-volume
%fields do not depend on given coordinate and hence the theory
%effectively reduces to D(p-1)-brane, for review see
%\cite{Simon:2011rw}. Clearly such a construction requires that
%Dp-brane is in the static gauge from the beginning  while in our
%case we want to fix the gauge at the end of given procedure.

Let us outline results derived in given paper. We generalize the
canonical description of T-duality transformations for the string
with no gauge fixing imposed. Then we show that the sequence of T
dualities and Wick rotation in target space-time together with the
uniform gauge fixing leads to the theory that is equivalent to the
double Wick rotated gauge fixed theory. On the other hand we show
that given procedure when applied to D1-brane action again gives
theory that is equivalent to the double Wick rotated uniform gauge
fixed theory which however cannot be interpreted as T-duality
transformations.

This paper is organized as follows. In the next section
(\ref{second}) we determine  T-duality transformations of the
background fields  as a
 canonical transformation of the Nambu-Gotto action for the
 fundamental string.
In section (\ref{third}) we perform sequence of T-duality
transformations and Wick rotation in the Hamiltonian formulation of
given theory and discuss its relation with the double Wick rotated
uniform gauge fixed action.
 Section (\ref{fourth}) is devoted to the
generalization of given procedure to the case of D1-brane and
finally in section (\ref{fifth}) we perform the combinations of
T-duality transformations and Wick rotation in case of  Hamiltonian
formulation of D1-brane theory.

\section{T-duality for Fundamental String in Canonical
Formalism}\label{second}
 We would like to perform the analysis of
T-duality as the canonical transformation of the NG action,
following \cite{Alvarez:1994wj}.
%\begin{eqnarray}
%S=-\frac{T_{D1}}{2}\int d\tau d\sigma [ \gamma^{\alpha\beta} (
%g_{11}\partial_\alpha \phi^1 \partial_\beta
%\phi^1+2g_{1\mu}\partial_\alpha \phi^1\partial_\beta
%x^\mu)+\nonumber
%\\
%+\epsilon^{\alpha\beta} b_{1\mu}
%\partial_\alpha \phi^1\partial_\beta x^\mu+
%b_{\mu 1}\partial_\alpha x^\mu\partial_\beta \phi^1
% \nonumber \\
%\end{eqnarray}
We start with the Nambu-Gotto action for the fundamental string
\begin{equation}\label{SNG}
S_{NG}=-\frac{1}{2\pi\alpha'}\int d\tau  d\sigma \left[ \sqrt{-\det
g} +\frac{1}{2}\epsilon^{\alpha\beta}B_{MN}\partial_\alpha
x^M\partial_\beta x^N\right] \ ,
\end{equation}
where $g_{\alpha\beta}=G_{MN}\partial_\alpha x^M
\partial_\beta x^N \ , \epsilon^{\tau\sigma}=-\epsilon^{\sigma\tau}=1$ and where $x^M,
M=0,\dots,d$ label the embedding coordinates of the string. Further,
$G_{MN}(x),B_{MN}(x)$ are components of the background gravitational
and NS-NS fields respectively.

As the first step we proceed to the Hamiltonian formulation of the
action (\ref{SNG}). From (\ref{SNG}) we obtain the conjugate momenta
\begin{eqnarray}\label{pMNG}
p_M=-\frac{1}{2\pi\alpha'}G_{MN}\partial_\alpha x^N g^{\alpha\tau}
\sqrt{-\det g }-\frac{1}{2\pi\alpha'}B_{MN}\partial_\sigma x^N \ .
\nonumber \\
\end{eqnarray}
If we now define
$\Pi_M=p_M+\frac{1}{2\pi\alpha'}B_{MN}\partial_\sigma x^N$ we obtain
from (\ref{pMNG}) following primary  constraint
\begin{eqnarray}\label{mHtau}
\mH_\tau=(2\pi\alpha')\Pi_M
G^{MN}\Pi_N+\frac{1}{(2\pi\alpha')}G_{MN}\partial_\sigma x^M
\partial_\sigma x^N=0
\nonumber \\
\end{eqnarray}
together with the spatial diffeomorphism constraint
\begin{equation}\label{mHsigma}
\mH_\sigma=p_M\partial_\sigma x^M \ .
\end{equation}
Then it can be shown that the bare Hamiltonian density defined as
$H_B=p_M\partial_\tau x^M-\mL$ vanishes identically and the extended
Hamiltonian density is a sum of the primary constraints of the
theory
\begin{equation}
\mH=\lambda_\tau \mH_\tau+\lambda_\sigma \mH_\sigma \ ,
\end{equation}
where $\lambda_\tau,\lambda_\sigma$ are Lagrange multipliers
corresponding to the constraints $\mH_\tau\approx 0 \ , \mH_\sigma
\approx 0$. It can be further shown that $\mH_\tau,\mH_\sigma$ are
the first class constraints that are generators of two dimensional
diffeomorphism of the world-sheet.

 Let us presume that
there is a direction in the target space-time  that is invariant
under constant shift
\begin{equation}
\theta\rightarrow \theta+\epsilon \ , \quad \epsilon=\mathrm{const}
\ .
\end{equation}
In other words all background fields do not depend on $\theta$. Our
goal is to perform the canonical transformation from $\theta$ to
$\ttheta$. Let us presume that this generating function has the form
\begin{equation}\label{defG}
G(\theta,\ttheta)=\frac{1}{4\pi\alpha'}\int d\sigma (
\partial_\sigma \theta \ttheta-\theta \partial_\sigma\ttheta) \ ,
\end{equation}
where we presume that $\theta$ has canonical dimension
$[\theta]=\mathrm{length}$.  Let us denote the momentum conjugate to
$\ttheta$ as $p_{\ttheta}$. From the definition of the canonical
transformations we derive following relations between canonical
momenta $p_\theta$ and $p_{\ttheta}$
% Now we have following relation between
%different Lagrangians
%\begin{eqnarray}
%p_\theta \partial_\tau
%\theta-H(p_\theta,\theta)=p_{\ttheta}\partial_\tau
%\ttheta-K(p_{\ttheta},\ttheta)+
%+\frac{d G(\theta,p_\theta)}{d\tau}=\nonumber \\
%p_{\ttheta}\partial_\tau \ttheta-K(p_{\ttheta},\ttheta)+
%+\frac{\partial G}{\partial t}+ \int_0^l d\sigma (\frac{\delta
%G}{\delta \theta(\sigma)}\partial_\tau\theta(\sigma)+ \frac{\delta
%G}{\delta \ttheta(\sigma)}
%\partial_\tau \ttheta(\sigma))
%=\nonumber \\
%\end{eqnarray}
%and hence
%\begin{equation}
%p_{\ttheta}=-\frac{\delta G}{\delta \ttheta} \ , \quad
%p_\theta=\frac{\delta G}{\delta \theta} \
%%\quad
%%K(p_{\ttheta},\ttheta)=H(p_\theta,\theta)+\frac{\partial G}{\partial
%%t} \ .
%\end{equation}
\begin{eqnarray}
p_{\ttheta}&=&-\frac{\delta G}{\delta
\ttheta}=-\frac{1}{2\pi\alpha'}\partial_\sigma \theta  \ , \nonumber
\\
 p_\theta&=&\frac{\delta G}{\delta \theta}= -\frac{1}{2\pi\alpha'}\partial_\sigma \ttheta \ .
\nonumber \\
\end{eqnarray}
Now we obtain canonically dual  Hamiltonian when we replace
$\partial_\sigma \theta$ with $-(2\pi\alpha') p_{\ttheta}$ and
$p_\theta$ with $-\frac{1}{2\pi\alpha'}\partial_\sigma \ttheta$ in
$\mH_\tau$ and $\mH_\sigma$ given in (\ref{mHtau}) and
(\ref{mHsigma}).
 Explicitly, we find
\begin{eqnarray}\label{mHtaudual}
 \tmH_\tau&=&
(2\pi\alpha') \hPi_\mu G^{\mu\nu}\hPi_\nu-2(2\pi\alpha') \hPi_\mu
G^{\mu\nu}B_{\nu\theta} p_{\ttheta}+ (2\pi\alpha')
p_{\ttheta}B_{\mu\theta}G^{\mu\nu}B_{\nu\theta}p_{\ttheta}-\nonumber
\\
&-&2\hPi_\mu G^{\mu\theta}\partial_\sigma \ttheta+2 \hPi_\mu
G^{\mu\theta} B_{\theta \nu}\partial_\sigma x^\nu +
2B_{\mu\theta}p_{\ttheta}G^{\mu\theta}
\partial_\sigma \ttheta-
2B_{\mu\theta} p_{\theta} G^{\mu\theta}
B_{\theta\nu}\partial_\sigma x^\nu+\nonumber \\
&+&\frac{1}{2\pi\alpha'}G^{\theta\theta}(\partial_\sigma
\ttheta)^2-2\frac{1}{2\pi\alpha'}
\partial_\sigma \ttheta G^{\theta\theta}B_{\theta\mu}\partial_\sigma
x^\mu+ \frac{1}{(2\pi\alpha')} B_{\theta\mu}\partial_\sigma x^\mu
G^{\theta\theta}B_{\theta\nu}\partial_\sigma x^\nu+\nonumber \\
 &+&(2\pi\alpha')G_{\theta\theta}
p_{\ttheta}^2-2 G_{\mu \theta}\partial_\sigma x^\mu
p_{\ttheta}+\frac{1}{(2\pi\alpha')}G_{\mu\nu}\partial_\sigma x^\mu
\partial_\sigma x^\nu \ , \nonumber \\
\tmH_\sigma&=& p_\mu\partial_\sigma x^\mu+p_{\ttheta}\partial_\sigma
\ttheta \ , \nonumber \\
\end{eqnarray}
where
\begin{equation}
\hPi_\mu=p_\mu+\frac{1}{2\pi\alpha'}B_{\mu\nu}\partial_\sigma x^\nu
\ ,
\end{equation}
$\mu,\nu=0,2,\dots,d$  . We see from  (\ref{mHtaudual}) that it is
very difficult to identify the theory in dual picture. To do this it
is more instructive to proceed to the Lagrangian formulation of the
theory. Explicitly, with the help of (\ref{mHtaudual}) we derive
following relations
\begin{eqnarray}
\partial_\tau x^\mu&=&\pb{x^\mu,H}=
2\lambda_\tau[(2\pi\alpha')G^{\mu\nu}\hPi_\nu-(2\pi\alpha')
G^{\mu\nu}B_{\nu\theta}p_{\ttheta}-G^{\mu\theta}\partial_\sigma
\ttheta+G^{\mu \theta}B_{\theta\nu}\partial_\sigma
x^\nu]+\lambda_\sigma \partial_\sigma x^\mu \ , \nonumber \\
\partial_\tau \ttheta&=&\pb{\ttheta,H}=
2\lambda_\tau[-(2\pi\alpha')\Pi_\mu
G^{\mu\nu}B_{\nu\theta}+(2\pi\alpha')B_{\mu\theta}G^{\mu\nu}B_{\nu\theta}
p_{\ttheta}+B_{\mu\theta}G^{\mu\theta}\partial_\sigma
\ttheta-\nonumber \\
&-&B_{\mu\theta}G^{\mu\theta}B_{\theta\nu}\partial_\sigma
x^\nu+(2\pi\alpha')G_{\theta\theta}p_{\ttheta}- G_{\theta
\nu}\partial_\sigma x^\nu]+\lambda_\sigma
\partial_\sigma \ttheta \ . \nonumber \\
\end{eqnarray}
Then after some algebra we find
\begin{eqnarray}\label{gPi}
g^{\mu\nu}\hPi_\nu&=&\frac{1}{2(2\pi\alpha')\lambda_\tau}(\bX^\mu+2\lambda_\tau
\bV^\mu+ 2(2\pi\alpha')\lambda_\tau
G^{\mu\nu}B_{\nu\theta}p_{\ttheta}) \ , \nonumber \\
p_{\ttheta}&=&\frac{1}{2(2\pi\alpha')\lambda_\tau G_{\theta\theta}}(
\Theta+ B_{\mu\theta}\bX^\mu
% - \lambda_\sigma \partial_\sigma x^\mu+
+2\lambda_\tau G_{\mu\theta}\partial_\sigma x^\mu) \ ,
\nonumber \\
\end{eqnarray}
where
\begin{equation}\label{defXmu}
\bV^\mu=G^{\mu\theta}\partial_\sigma \ttheta- G^{\mu\theta}
B_{\theta\nu}\partial_\sigma x^\nu \ , \quad \bX^\mu=\partial_\tau
x^\mu-\lambda_\sigma \partial_\sigma x^\mu \ , \quad
\Theta=\partial_\tau \ttheta-\lambda_\sigma \partial_\sigma \ttheta
\ .
\end{equation}
In order to express $\hPi_\mu$ from (\ref{gPi}) we have to find the
inverse matrix to $G^{\mu\nu}$. Recall that  by definition
\begin{equation}\label{defgMN}
G_{MN}G^{NK}=\delta_M^K \ , \quad G_{\mu M}G^{N\nu}= G_{\mu
\rho}G^{\rho N}+G_{\mu\theta}G^{\theta\nu}=\delta_\mu^\nu
\end{equation}
and hence we see that $G_{\mu\nu}$ is not inverse to $G^{\mu\nu}$.
It turns  out that given matrix has the form
\begin{equation}
h_{\mu\nu}=G_{\mu\nu}-\frac{G_{\mu
\theta}G_{\nu\theta}}{G_{\theta\theta}}
\end{equation}
as can be easily seen from (\ref{defgMN})
\begin{eqnarray}
h_{\mu\nu}G^{\nu\rho}
%= g_{\mu\nu}g^{\nu\rho}- \frac{g_{\mu
%\theta}g_{\nu\theta}}{g_{\theta\theta}}g^{\nu\rho}= \nonumber \\
%=\delta_\mu^\rho-g_{\mu\theta}g^{\theta\rho}+\frac{g_{\mu\theta}}{g_{\theta\theta}}
%g_{\theta\theta}g^{\theta\rho}
=\delta_\mu^\rho \ , \quad
G^{\theta\mu}h_{\mu\nu}=-\frac{G_{\theta\nu}}{G_{\theta\theta}} \ .
\nonumber \\
\end{eqnarray}
%using
%\begin{eqnarray}
%g_{\mu\nu}g^{\nu\rho}=\delta_\mu^\rho-g_{\mu\theta}g^{\theta\rho} \
%, g_{\theta\nu}g^{\nu\rho}+g_{\theta\theta}g^{\theta\rho}=0
%\nonumber \\
%g^{\theta\mu}g_{\mu\theta}=1-g^{\theta\theta}g_{\theta\theta} \ .
%\nonumber \\
%\end{eqnarray}
Then we obtain
\begin{equation}
\hPi_\mu=\frac{1}{2(2\pi\alpha')\lambda_\tau}h_{\mu\nu}
(\bX^\nu+2\lambda_\tau \bV^\nu+ 2(2\pi\alpha')\lambda_\tau
G^{\nu\rho}B_{\rho\theta}p_{\ttheta}) \
\end{equation}
and after some algebra we find the Lagrangian for dual theory in the
form
\begin{eqnarray}\label{LTdual}
\tilde{\mL}&=&p_{\ttheta}\partial_\tau \ttheta+p_\mu\partial_\sigma
x^\mu-
\lambda_\tau \tmH_\tau-\lambda_\sigma \tmH_\sigma= \nonumber \\
&=&\frac{1}{4(2\pi\alpha')\lambda_\tau}\left(\tg_{\tau\tau}+2\lambda_\sigma
\tg_{\tau\sigma}+\lambda_{\sigma}^2
\tg_{\sigma\sigma}\right)-\frac{1}{2\pi\alpha'}
\lambda_\tau \tg_{\sigma\sigma} -\nonumber \\
&-&\frac{1}{2\pi\alpha'}\partial_\tau x^\mu
\tB_{\mu\nu}\partial_\sigma x^\nu
%-\frac{1}{2\pi\alpha'}\left(\frac{g_{\mu\theta}}{g_{\theta\theta}}
%\partial_\tau x^\mu \partial_\sigma \ttheta-
%\frac{g_{\mu\theta}}{g_{\theta\theta}}\partial_\tau \ttheta
%\partial_\sigma
%x^\mu\right)
 -\frac{1}{2\pi\alpha'}\partial_\tau \ttheta
\tB_{\ttheta\mu}\partial_\sigma x^\mu- \frac{1}{2\pi\alpha'}
\partial_\tau x^\mu \tB_{\mu\ttheta}\partial_\sigma \ttheta
 \ ,  \nonumber \\
\end{eqnarray}
where
\begin{eqnarray}
 \tg_{\alpha\beta}=\tG_{\ttheta\ttheta}\partial_\alpha
\ttheta\partial_\beta \ttheta+ \tG_{\ttheta\mu}\partial_\alpha
\ttheta\partial_\beta x^\mu+\tG_{\mu \ttheta}\partial_\alpha x^\mu
\partial_\beta \ttheta+\tG_{\mu\nu}\partial_\alpha x^\mu
\partial_\beta x^\nu \ ,
\nonumber \\
\end{eqnarray}
and where we defined T-dual components of the metric and NS-NS two
form
\begin{eqnarray}
\tG_{\mu\nu}&=&h_{\mu\nu}+\frac{B_{\mu\theta}B_{\nu\theta}}{G_{\theta\theta}}
\ ,
\tG_{\mu\ttheta}=\tG_{\ttheta\mu}=\frac{B_{\mu\theta}}{G_{\theta\theta}}
\ , \tG_{\ttheta\ttheta}=\frac{1}{G_{\theta\theta}}\nonumber \\
\tB_{\mu\nu}&=& B_{\mu\nu}+\frac{B_{\mu\theta}G_{\theta
\nu}}{G_{\theta\theta}} -\frac{G_{\mu\theta} B_{\theta
\nu}}{G_{\theta\theta}} \ , \quad
\tB_{\ttheta\mu}=-\tB_{\mu\ttheta}=-\frac{G_{\theta\mu}}{G_{\theta\theta}}
\nonumber
\\
\end{eqnarray}
that coincide with Buscher's transformations
\cite{Buscher:1987sk,Buscher:1987qj}.
In order to have more familiar form of the Lagrangian density we
solve  the equations of motion for $\lambda_\tau$ and
$\lambda_\sigma$ that follow from (\ref{LTdual}).  Explicitly, the
equation of motion for $\lambda_\sigma$ has the form
\begin{equation}\label{eqlambdasigma}
\tg_{\tau\sigma}+\lambda_\sigma \tg_{\sigma\sigma}=0
\end{equation}
while the equation of motion for $\lambda_\tau$ takes the form
\begin{equation}
-\frac{1}{4\lambda^2_\tau}[\tg_{\tau\tau}+2\lambda_\sigma
\tg_{\tau\sigma}+\lambda_{\sigma}^2
\tg_{\sigma\sigma}]+\tg_{\sigma\sigma}=0
\end{equation}
that together with (\ref{eqlambdasigma}) implies
\begin{equation}
\lambda_\tau=\frac{\sqrt{\tg_{\tau\sigma}^2-
\tg_{\tau\tau}\tg_{\sigma\sigma}}}{2\tg_{\sigma\sigma}} \ .
\end{equation}
Inserting these expressions into the Lagrangian density
(\ref{LTdual}) we obtain the final result
\begin{equation}
\tilde{\mL}=-\frac{1}{2\pi\alpha'}\left(\sqrt{-\det \tg}
+\frac{1}{2}\epsilon^{\alpha\beta}\tb_{\alpha\beta}\right)
\end{equation}
that is the Lagrangian density for Nambu-Gotto string in T-dual
background.
%Now
%the equation of motion for $\lambda_\tau,\lambda_\sigma $have
%the form
%\begin{eqnarray}
%\frac{1}{4\lambda_\tau}[\bA_2+2\lambda_\sigma \bA_3]+\bB=0
%\Rightarrow \lambda_\sigma=-\frac{4\bB \lambda_\tau+\bA_2}{2\bA_3} \
%, \nonumber \\
%-\frac{1}{4\lambda_\tau^2}[\bA_1+\bA_2\lambda_\sigma+\bA_3\lambda_\sigma^2]+
%\bC=0 \nonumber \\
%\end{eqnarray}
%Now inserting the value $\lambda_\sigma$ to the equation of motion
%for $\lambda_\tau$ we obtain
%\begin{equation}
%\lambda_\tau=\sqrt{\frac{4\bA_1\bA_3-\bA_2^2}{16(\bC\bA_3-\bB^2)}}
%\end{equation}
%Finally inserting these results into the Lagrangian density we
%obtainevenlyexitif
%\begin{eqnarray}
%\mL=\sqrt{\frac{\bC \bA_3-\bB^2}{4\bA_1\bA_3-\bA_2^2}}
%[\bA_1-\frac{\bA_2^2}{4\bA_3}]+\frac{\bB^2}{\bA_3}\lambda_\tau-
%\nonumber \\
%-\frac{4\bB^2 \lambda_\tau+\bA_2\bB^2}{2\bA_3}+\lambda_\tau \bC
%-\frac{1}{2\pi\alpha'}
%\partial_\tau x^\mu b_{\mu\nu}\partial_\sigma
%x^\nu+\frac{g_{\mu\theta}}{g_{\theta\theta}}\partial_\sigma x^\mu
%\partial_\tau \theta \nonumber \\
%=\frac{1}{2}\sqrt{\frac{4\bA_1\bA_3-\bA_2^2}{\bA_3^2}(\bC \bA_3-\bB^2)}
%\end{eqnarray}

\section{T-duality and Double Wick rotation}\label{third}
We have shown in the previous section that canonical transformation
in the Hamiltonian formulation of NG string gives the action for the
string in T-dual background. Using this result we focus in this
section on the relation between T-duality and double Wick rotation.
We presume that the background has the form
\cite{Arutyunov:2013ega,Arutyunov:2014ota,Arutyunov:2014cra}
\begin{eqnarray}\label{lightconemetric}
ds^2&=&G_{MN}dx^M
dx^N=G_{tt}dt^2+G_{\varphi\varphi}d\varphi^2+G_{\mu\nu}dx^\mu dx^\nu \ , \nonumber \\
 B&=&B_{MN}dX^M dX^N=B_{\mu\nu}dx^\mu dx^\nu \ , \nonumber \\
\end{eqnarray}
where $\mu,\nu$ denote the transverse directions.
 Following
\cite{Arutyunov:2013ega,Arutyunov:2014ota} we introduce light cone
coordinates
\begin{equation}\label{defxplus}
x^-=\varphi-t \ , \quad x^+=(1-a)t+a\varphi \
\end{equation}
with inverse relations
\begin{equation}
t=x^+-ax^- \ , \quad \varphi=x^++(1-a)x^- \ .
\end{equation}
Then corresponding metric components have the form
%In terms of these variables the line element has the form
%\begin{eqnarray}
%ds^2=G_{\varphi\varphi }d\varphi^2+G_{tt}dt^2=
%(G_{tt}+G_{\varphi\varphi})(dx^+)^2+\nonumber
%\\
%+2(-aG_{tt}+(1-a)G_{\varphi\varphi})
%dx^+dx^-+(G_{tt}a^2+(1-a)^2G_{\varphi\varphi})(dx^-)^2 \nonumber \\
%\end{eqnarray}
%so that we have relation between metric components
\begin{equation}\label{Gplusplus}
G_{++}=G_{tt}+G_{\varphi\varphi} \ , \quad
G_{--}=G_{tt}a^2+(1-a)^2G_{\varphi\varphi} \ , \quad  G_{+-}= -a
G_{tt}+(1-a)G_{\varphi\varphi} \
\end{equation}
with inverse
\begin{eqnarray}\label{Gplusplusin}
G^{++}&=&
%\frac{G_{--}}{G_{++}G_{--}-G_{+-}G_{+-}}=
\frac{G_{tt}a^2+(1-a)^2G_{\varphi\varphi}}{G_{tt}G_{\varphi\varphi}}
\ ,  \quad  G^{--}=
% \frac{G_{++}}{G_{++}G_{--}-G_{+-}G_{+-}}
\frac{G_{tt}+G_{\varphi
 \varphi}}{G_{tt}G_{\varphi\varphi}} \ ,
 \nonumber \\
 G^{+-}&=&
 %-\frac{G_{+-}}{G_{++}G_{--}-G_{+-}G_{+-}}=
 \frac{aG_{tt}-(1-a)G_{\varphi\varphi}}{G_{tt}G_{\varphi\varphi}} \
 .
 \nonumber \\
\end{eqnarray}
In the light cone coordinates the  Hamiltonian and diffeomorphism
constraints have the form
\begin{eqnarray}\label{mHtaustring}
\mH_\tau&=&(2\pi\alpha')p_+
G^{++}p_++2(2\pi\alpha')p_+G^{+-}p_-+(2\pi\alpha')p_-G^{--}p_-+
\nonumber \\
&+& \frac{1}{2\pi\alpha'}[G_{++}(\partial_\sigma x^+)^2+2G_{+-}
\partial_\sigma x^+\partial_\sigma x^-+G_{--}(\partial_\sigma
x^-)^2]+\mH_x \ , \nonumber \\
\mH_\sigma&=&p_+\partial_\sigma x^++p_-\partial_\sigma x^-+p_\mu
\partial_\sigma x^\mu \ , \nonumber \\
\end{eqnarray}
where
\begin{eqnarray}
 \mH_x=(2\pi\alpha')\Pi_\mu G^{\mu\nu}\Pi_\nu+\frac{1}{2\pi\alpha'}
G_{\mu\nu}\partial_\sigma x^\mu \partial_\sigma x^\nu  \ , \quad
\Pi_\mu=p_\mu+\frac{1}{2\pi\alpha'}B_{\mu\nu}\partial_\sigma x^\nu \
. \nonumber
\\
\end{eqnarray}
Now we are ready to study the relation between T-duality and double
Wick rotation. Let us presume that the background does not depend on
$x^-$. Our goal is to perform the canonical transformation from
$x^-$ to $\psi$, where, following discussion presented in previous
section,  the generating function has the form
\begin{equation}
G(\theta,\psi)=\frac{1}{4\pi\alpha'}\int d\sigma (
\partial_\sigma x^- \psi-x^- \partial_\sigma \psi) \ .
\end{equation}
%where we presume that $\psi$ has dimension $[\psi]=\mathrm{length}$
%and we demand that $G$ should be dimensionless.
Let us denote the
momentum conjugate to $\psi$ as $p_{\psi}$. Then we obtain
\begin{eqnarray}
p_{\psi}=-\frac{1}{2\pi\alpha'}\partial_\sigma x^-  \ , \quad p_{-}=
-\frac{1}{2\pi\alpha'}\partial_\sigma \psi \
\nonumber \\
\end{eqnarray}
so that we  obtain T-dual Hamiltonian when we replace
$\partial_\sigma x^-$ with $-(2\pi\alpha')p_{\psi}$ and $p_-$ with
$-\frac{1}{2\pi\alpha'}\partial_\sigma \psi$ and hence
\begin{eqnarray}\label{mHtaustringTdual}
\tmH_\tau&=& (2\pi\alpha')
p_+G^{++}p_+-2p_+G^{+-}\partial_\sigma\psi
+\frac{1}{2\pi\alpha'}G^{--}(\partial_\sigma
\psi)^2+\frac{1}{2\pi\alpha'}
\left(G_{++}(\partial_\sigma x^+)^2+\right.\nonumber \\
&-& \left. 4\pi\alpha' G_{+-}\partial_\sigma x^+
p_{\psi}+(2\pi\alpha')^2 G_{--}(p_\psi)^2\right)+ \mH_x \ ,
 \nonumber \\
\tmH_\sigma &=&p_+\partial_\sigma x^++p_\psi\partial_\sigma
\psi+p_\mu
\partial_\sigma x^\mu \ .  \nonumber \\
\end{eqnarray}
Then following \cite{Arutyunov:2014cra} we perform analytic
continuation in the target space-time
\begin{equation}\label{analcon}
(x^+,\psi)\rightarrow (i\tpsi, -i\tx^+)
 \ , \quad (p_+,p_\psi)\rightarrow (-i\tp_\psi, i\tp_+) \ .
 \end{equation}
so that we obtain
\begin{eqnarray}
\tmH_\tau&=& -(2\pi\alpha')\tp_\psi G^{++}\tp_\psi+2 \tp_\psi
G^{+-}\partial_\sigma \tx^+
-\frac{1}{2\pi\alpha'}G^{--}(\partial_\sigma \tx^+)^2 +\nonumber \\
&+&\frac{1}{2\pi\alpha'} \left(-G_{++}(\partial_\sigma \tpsi)^2+
4\pi\alpha' G_{+-}\partial_\sigma \tpsi
\tp_{+}-(2\pi\alpha')^2G_{--}(\tp_+)^2\right)+ \mH_x \ ,
 \nonumber \\
\tmH_\sigma &=&p_+\partial_\sigma x^++\tp_\psi\partial_\sigma
\psi+p_\mu
\partial_\sigma x^\mu \ .  \nonumber \\
\end{eqnarray}
Note that the way how the conjugate momenta $p_+,p_\psi$ transform
under the  analytic continuation  is given by the requirement that
all terms in $\tmH_\sigma$ come with $+$ sign since we demand that
the string theory is invariant under  world-sheet diffeomorphism and
hence Wick rotated $\tmH_\sigma$ should have the same form as the
original one.

Finally we perform T-duality transformation along $\tpsi$ direction
that gives
\begin{equation}
\tp_{\psi}=-\frac{1}{2\pi\alpha'}\partial_\sigma \tphi \ , \quad
p_{\tphi}=-\frac{1}{2\pi\alpha'}\partial_\sigma \tpsi \
\end{equation}
so that we obtain the final form of the Hamiltonian and
diffeomorphism constraints
\begin{eqnarray}
\tmH_\tau&=& -\frac{1}{2\pi\alpha'}(\partial_\sigma \tphi)^2 G^{++}
-\frac{2}{2\pi\alpha'}\partial_\sigma \tphi G^{+-}\partial_\sigma
\tx^+
-\frac{1}{2\pi\alpha'}G^{--}(\partial_\sigma \tx^+)^2 +\nonumber \\
&+&(2\pi\alpha') \left(-G_{++}p_{\tphi}^2- 2G_{+-}p_{\tphi}
\tp_{+}-G_{--}(\tp_+)^2\right)+ \mH_x \ ,
 \nonumber \\
\tmH_\sigma &=&p_+\partial_\sigma x^++\tp_\psi\partial_\sigma
\psi+p_\mu
\partial_\sigma x^\mu \ .  \nonumber \\
\end{eqnarray}
In order to find the Hamiltonian for the physical degrees of freedom
we have to fix the gauge. It turns out that it is natural to use
uniform gauge fixing
%Finally we fix the gauge by imposing
%\begin{equation}
%\tphi=\sigma \ , x^+=\tau \ .
%\end{equation}
%so that from $\mH_\sigma=0$ we obtain
%\begin{equation}
%p_{\tphi}=-p_\mu\partial_\sigma x^\mu
%\end{equation}
% so that $\mH_\tau$ is equal
%to
%\begin{eqnarray}
%\mH_\tau&=& -\frac{1}{2\pi\alpha'}(\partial_\sigma \tphi)^2 G^{++}
%-\frac{2}{2\pi\alpha'}\partial_\sigma \tphi G^{+-}\partial_\sigma
%\tx^+
%-\frac{1}{2\pi\alpha'}G^{--}(\partial_\sigma \tx^+)^2 +\nonumber \\
%&+&(2\pi\alpha')
%\left(-G_{++}p_{\tphi}^2-\right.\nonumber \\
%&-& \left. 2G_{+-}p_{\tphi} \tp_{+}-G_{--}(\tp_+)^2\right)+ \mH_x \
%,
% \nonumber \\
%\mH_\sigma &=&p_+\partial_\sigma x^++\tp_\psi\partial_\sigma
%\psi+p_\mu
%\partial_\sigma x^\mu \ ,  \nonumber \\
%\end{eqnarray}
\begin{equation}
p_{\tphi}=\frac{1}{2\pi\alpha'} \ , \quad  x^+=\tau \ .
\end{equation}
Then from $\tmH_\sigma=0$ we find
\begin{equation}
\partial_\sigma \tphi=-(2\pi\alpha')(p_\mu\partial_\sigma x^\mu)
\end{equation}
and hence the Hamiltonian constraint is equal to
\begin{eqnarray}\label{mHtaufixed}
\tmH_\tau&=& -(2\pi\alpha')(\partial_\sigma x^\mu p_\mu)^2 G^{++}
-\frac{1}{2\pi\alpha'}G_{++}- 2G_{+-} \tp_{+}-(2\pi\alpha')
\tp_+G_{--}\tp_++ \mH_x=0 \ .
 \nonumber \\
\end{eqnarray}
Note that due to the gauge fixing the constraints
$\tmH_\tau,\tmH_\sigma$ vanish strongly and hence (\ref{mHtaufixed})
serves as the quadratic equation for $\tp_+$. In fact, $-\tp_+$
should be identified as the Hamiltonian density for the physical
degrees of freedom after gauge fixing.

Now we would like to compare the equation (\ref{mHtaufixed}) with
the equation that defines the Hamiltonian density  for the physical
degrees of freedom for of the uniform gauge fixed  string. Note that
this gauge is imposed in the Hamiltonian formulation of the string
we identify $p_-=\frac{1}{2\pi\alpha'}$. Equivalently we can impose
given gauge in T-dual theory when we identify $\psi$ with $\sigma$
\cite{Kruczenski:2004cn}. This construction has an advantage   since
it does not require to go to the Hamiltonian formulation of given
theory which could be extremely difficult in case of Green-Schwarz
action. However in our case we can either choose
 the Hamiltonian constraint (\ref{mHtaustring}) and impose the
gauge $p_-=\frac{1}{2\pi\alpha'},x^+=\tau$ or use T-dual Hamiltonian
constraint (\ref{mHtaustringTdual}) with the following gauge fixing
functions
\begin{equation}\label{Tdualfix}
\psi=\sigma \ , x^+=\tau \ .
\end{equation}
We choose the second possibility and using (\ref{Tdualfix}) in
$\mH_\sigma$ we obtain $p_\psi=-(\partial_\sigma x^\mu p_\mu)$ that
together with (\ref{Tdualfix}) implies that (\ref{mHtaustringTdual})
has the form
\begin{eqnarray}\label{HfixTdual}
\mH_\tau= (2\pi\alpha')p_+G^{++}p_+-2p_+G^{+-}
+\frac{1}{2\pi\alpha'} G^{--} +(2\pi\alpha')
G_{--}(p_\mu\partial_\sigma x^\mu)^2+ \mH_x=0 \ .
 \nonumber \\
\end{eqnarray}
This is quadratic equation for $p_+$. Comparing (\ref{HfixTdual})
with (\ref{mHtaufixed}) we see that they have the same form when we
define metric components
\begin{equation}
\tG^{++}=-G_{--} \ , \quad \tG^{--}=-G_{++} \ , \quad \tG^{+-}=
G_{+-} \ .
\end{equation}
In other words, the sequence of T-dualities and analytic
continuation in target space-time  implies the transformation of the
components of the target metric that has the same form as the double
Wick rotation in the uniform gauge fixed bosonic string
\cite{Arutyunov:2014cra}. On the other hand we also see that
components of NS-NS two form that are transverse to the directions
where these dualities were performed do not transform. At this place
we see the difference with the double Wick rotation of the gauge
fixed action \cite{Arutyunov:2014cra} since in this case these
components change the sign. In order to resolve this issue we can
extend  the sequence of T-duality transformation and Wick rotation
with an additional transformation $B\rightarrow -B$ since as we
argued in the introduction whenever the background field $B$ is
solution of the supergravity equations of motion so $-B$ is too. In
other words we have the equivalence between world-sheet double Wick
rotation and the sequence: "T-duality-target space Wick
rotation-T-duality-$B\rightarrow -B$.

\section{D1-brane and Duality Transformation}\label{fourth}
In this section we apply the same ideas to the case of DBI and WZ
action for D1-brane.  Let us start with D1-brane action
\begin{eqnarray}\label{SDbrane}
S&=&-T_{D1}\int d\tau d\sigma e^{-\Phi} \sqrt{-\det
(g_{\alpha\beta}+b_{\alpha\beta}+
(2\pi\alpha')F_{\alpha\beta})}+\nonumber \\
&+& T_{D1}\int d\tau d\sigma[
C^{(0)}(b_{\tau\sigma}+(2\pi\alpha')F_{\tau\sigma})+C_{\tau\sigma}^{(2)}]
\ ,
\nonumber \\
\end{eqnarray}
where
\begin{equation}
g_{\alpha\beta}=G_{MN}\partial_\alpha x^M\partial_\beta x^N \ ,
\quad b_{\alpha\beta}=B_{MN}\partial_\alpha x^M\partial_\beta x^N \
, \quad  C^{(2)}_{\tau\sigma}=C_{MN}^{(2)}\partial_\tau
x^M\partial_\sigma x^N \ ,
\end{equation}
and where $x^M(\tau,\sigma)$ are embedding coordinates for D1-brane
in given background. Further,  $F_{\alpha\beta}=\partial_\alpha
A_\beta-
\partial_\beta A_\alpha$ is the field strength of the world-volume
gauge field $A_\alpha,\alpha=\tau,\sigma$. Finally $T_{D1}$ is
D1-brane tension $T_{D1}=\frac{1}{2\pi\alpha'}$.

Before we proceed to the Hamiltonian formulation of the action
(\ref{SDbrane}) it is useful to use following formula
\begin{equation}
\det (g_{\alpha\beta}+b_{\alpha\beta}+(2\pi\alpha')
F_{\alpha\beta})= \det g+(b_{\tau\sigma}+(2\pi\alpha')
F_{\tau\sigma})^2 \
\end{equation}
that holds in two dimensions only.  Then from the action
(\ref{SDbrane}) we find momenta conjugate to $x^M,A_\sigma$ and
$A_\tau$ respectively
\begin{eqnarray}
p_M&=&T_{D1} \frac{e^{-\Phi}}{\sqrt{-\det g
-((2\pi\alpha')F_{\tau\sigma}+
b_{\tau\sigma})^2}}\left(G_{MN}\partial_\alpha x^N g^{\alpha
\tau}\det
g+\right.\nonumber \\
&+&\left.((2\pi\alpha')F_{\tau\sigma}+b_{\tau\sigma})B_{MN}\partial_\sigma
x^N\right)+T_{D1}(C^{(0)}B_{MN}\partial_\sigma x^N+C^{(2)}_{MN}\partial_\sigma x^N) \ , \nonumber \\
\pi^\sigma&=&\frac{e^{-\Phi}T_{D1}(2\pi\alpha')((2\pi\alpha')F_{\tau\sigma}+
b_{\tau\sigma})}{\sqrt{-\det g -((2\pi\alpha')F_{\tau\sigma}+
b_{\tau\sigma})^2}}+T_{D1}(2\pi\alpha')C^{(0)}\ , \quad
\pi^\tau\approx 0 \ .  \nonumber \\
\end{eqnarray}
Using these relations we find that the bare Hamiltonian is equal to
\begin{eqnarray}
H_B=\int d\sigma (p_M\partial_\tau x^M+\pi^\sigma \partial_\tau
A_\sigma-\mL)= \int d\sigma \pi^\sigma\partial_\sigma A_\tau
\end{eqnarray}
while we have three primary constraints
\begin{eqnarray}
\mH_\sigma&\equiv& p_M\partial_\sigma x^M\approx 0 \ , \quad \pi^\tau\approx 0 \ ,  \nonumber \\
\mH_\tau &\equiv & (2\pi\alpha')\Pi_M
G^{MN}\Pi_N+\frac{1}{2\pi\alpha'}\left(e^{-2\Phi}+\left(\pi^\sigma-
C^{(0)}\right)^2\right)G_{MN}\partial_\sigma x^M\partial_\sigma x^N
\ ,
\nonumber \\
\end{eqnarray}
where
\begin{eqnarray}
\Pi_M&\equiv &
p_M-\frac{\pi^\sigma}{2\pi\alpha'}B_{MN}\partial_\sigma x^N
-\frac{1}{2\pi\alpha'}C^{(2)}_{MN}\partial_\sigma x^N \ ,
% \quad
%g_{\sigma\sigma}\equiv g_{MN}\partial_\sigma x^M\partial_\sigma x^N
%\
\nonumber \\
%= \nonumber \\
%&=&T_{D1} \frac{e^{-\Phi}}{\sqrt{-\det g
%-((2\pi\alpha')F_{\tau\sigma}+
%b_{\tau\sigma})^2}}g_{MN}\partial_\alpha x^N g^{\alpha \tau}\det g \
% \ ,
%\nonumber \\
\end{eqnarray}
and where  we used the fact that $T_{D1}=\frac{1}{2\pi\alpha'}$.
%and where
%\begin{equation}
%g_{\sigma\sigma}\equiv g_{MN}\partial_\sigma x^M\partial_\sigma x^N
%\ .
%\end{equation}
According to the standard treatment of the constraint systems we
introduce the extended Hamiltonian with all primary constraints
included
\begin{equation}
H=\int d\sigma (\lambda_\tau\mH_\tau+\lambda_\sigma
\mH_\sigma-A_\tau\partial_\sigma\pi^\sigma+v_\tau \pi^\tau) \ ,
\end{equation}
where $\lambda_{\tau,\sigma}$ and $v_\tau$ are Lagrange multipliers
corresponding to the constraints $\mH_\tau,\mH_\sigma $ and
$\pi^\tau$.

Now the requirement of the preservation of the primary constraint
$\pi^\tau\approx 0$ implies the secondary constraint
\begin{equation}
\mG=\partial_\sigma \pi^\sigma\approx 0 \ .
\end{equation}
Then it can be shown that $\mH_\tau,\mH_\sigma$ are the first class
constraints, for details, see \cite{Kluson:2014uaa}.

Now we can formally proceed to the discussion of the canonical
transformation as in the case of fundamental string.
 Let us presume that
there is a direction that is invariant under constant shift
\begin{equation}
\theta\rightarrow \theta+\epsilon \ , \quad \epsilon=\mathrm{const}
\ .
\end{equation}
Then we again consider the  generating function into the form
\begin{equation}\label{canD1}
G(\theta,\ttheta)=\frac{1}{4\pi\alpha'}\int d\sigma (
\partial_\sigma \theta \ttheta-\theta \partial_\sigma\ttheta)
\end{equation}
that implies the relation between momenta $p_\theta$ and
$p_{\ttheta}$ respectively
%\begin{eqnarray}
%p_\theta \partial_\tau
%\theta-H(p_\theta,\theta)=p_{\ttheta}\partial_\tau
%\ttheta-K(p_{\ttheta},\ttheta)+
%+\frac{d G(\theta,p_\theta)}{d\tau}=\nonumber \\
%p_{\ttheta}\partial_\tau \ttheta-K(p_{\ttheta},\ttheta)+
%+\frac{\partial G}{\partial t}+ \int_0^l d\sigma (\frac{\delta
%G}{\delta \theta(\sigma)}\partial_\tau\theta(\sigma)+ \frac{\delta
%G}{\delta \ttheta(\sigma)}
%\partial_\tau \ttheta(\sigma))
%=\nonumber \\
%\end{eqnarray}
%and hence
%\begin{equation}
%p_{\ttheta}=-\frac{\delta G}{\delta \ttheta} \ , \quad
%p_\theta=\frac{\delta G}{\delta \theta} \ , \quad
%K(p_{\ttheta},\ttheta)=H(p_\theta,\theta)+\frac{\partial G}{\partial
%t} \ .
%\end{equation}
%From the generating function we find
\begin{eqnarray}
p_{\ttheta}=-\frac{1}{2\pi\alpha'}\partial_\sigma \theta  \ , \quad
p_\theta= -\frac{1}{2\pi\alpha'}\partial_\sigma \ttheta \ .
\nonumber \\
\end{eqnarray}
Then we  obtain dual Hamiltonian when we replace $\partial_\sigma
\theta$ with $-(2\pi\alpha')p_{\ttheta}$ and $p_\theta$ with
$-\frac{1}{2\pi\alpha'}\partial_\sigma \ttheta$ so that
\begin{eqnarray}\label{tildemH}
\tilde{\mH}_{\tau}&=&(2\pi\alpha')\Pi_\theta
G^{\theta\theta}\Pi_\theta+ 2(2\pi\alpha')\Pi_\theta G^{\theta
\mu}\Pi_\mu+(2\pi\alpha')\Pi_\nu
G^{\mu\nu}\Pi_\nu+\nonumber \\
&+&\frac{1}{2\pi\alpha'}\left(e^{-2\Phi}+\left(\pi^\sigma
-C^{(0)}\right)^2\right)G_{\mu\nu}\partial_\sigma
x^\mu\partial_\sigma x^\nu-\nonumber \\
&-&2\left(e^{-2\Phi}+\left(\pi^\sigma -C^{(0)}\right)^2\right)G_{\mu
\theta}\partial_\sigma x^\mu
p_{\ttheta}+\nonumber \\
&+& (2\pi\alpha')\left(e^{-2\Phi}+ \left(\pi^\sigma
-C^{(0)}\right)^2\right)G_{\theta\theta}p_{\ttheta}p_{\ttheta}
 \ ,
\nonumber \\
\tmH_\sigma&=&p_\mu\partial_\sigma x^\mu+p_{\ttheta}\partial_\sigma
\ttheta \ . \nonumber \\
\end{eqnarray}
In order to see how the background fields transform under this
duality transformation we should find corresponding Lagrangian.
Before we proceed to this question we should stress one important
point. In principle the electric flux that is given in
(\ref{tildemH}) is off-shell.
 However we know that this electric
flux is proportional to the number of the fundamental strings. Our
goal is to compare actions where this number is the same so that we
consider canonical transformations for D1-brane theory where we fix
the gauge symmetry so that $\pi^\sigma=\mathrm{const}$. In fact, if
$\pi^\sigma$ were the dynamical variable we would get very
complicated form of the Lagrangian density due to the fact that now
the Hamiltonian contains term like  $(\pi^\sigma)^2 p_{\ttheta}^2$.

In order to proceed to the Lagrangian formulation we introduce
following notations
\begin{eqnarray}
\Pi_\mu&=&
%p_\mu-\frac{\pi^\sigma}{(2\pi\alpha')}b_{\mu\nu}\partial_\sigma
%x^\nu+\frac{\pi^\sigma}{(2\pi\alpha')}b_{\mu\theta} p_{\ttheta} -\nonumber \\
% -T_{D1}C^{(2)}_{\mu\nu}\partial_\sigma x^\nu+
%T_{D1}C^{(2)}_{\mu\theta}p_{\ttheta}=
\hPi_\mu +\bV_\mu p_{\ttheta} \
, \quad  \bV_\mu=\pi^\sigma B_{\mu\theta}+C^{(2)}_{\mu\theta} \ , \nonumber \\
\hPi_\mu&=&
p_\mu-\frac{\pi^\sigma}{2\pi\alpha'}B_{\mu\nu}\partial_\sigma x^\nu
-\frac{1}{2\pi\alpha'}C^{(2)}_{\mu\nu}\partial_\sigma x^\nu \ ,
\quad
 \nonumber \\
 \Pi_\theta&=&-\frac{1}{2\pi\alpha'}(\partial_\sigma \ttheta +\bV_\mu \partial_\sigma
 x^\mu)
 \ , \quad
\bX=e^{-2\Phi}+(\pi^\sigma-C^{(0)})^2 \ . \nonumber \\
\end{eqnarray}
%\begin{eqnarray}
%\Pi_\theta g^{\theta\theta}\Pi_\theta+ 2\Pi_\theta g^{\theta
%\mu}\Pi_\mu+\Pi_\nu g^{\mu\nu}\Pi_\nu=\nonumber \\
%g^{\theta\theta}(\partial_\sigma \ttheta)^2 -2\bV_\mu
%\partial_\sigma x^\mu g^{\theta\theta}\partial_\sigma \ttheta+
%\partial_\sigma x^\mu (\bV_\mu
%g^{\theta\theta}\bV_\nu)\partial_\sigma x^\nu-\nonumber \\
%-2\partial_\sigma \ttheta g^{\theta\mu}\hPi_\mu-2\partial_\sigma
%\ttheta g^{\theta\mu}\bV_\mu p_{\ttheta}- 2\bV_\nu \partial_\sigma
%x^\nu g^{\theta\mu}\hPi_\mu-2\bV_\nu \partial_\sigma x^\nu
%g^{\theta\mu}\bV_\mu p_{\ttheta}+\nonumber \\
%+\hPi_\mu g^{\mu\nu}\hPi_\nu+2\hPi_\mu g^{\mu\nu}\bV_\nu
%p_{\ttheta}+\bV_\mu g^{\mu\nu}\bV_\nu p_{\ttheta}p_{\ttheta} \ .
%\nonumber \\
%\end{eqnarray}
so that  we can write $\tmH_\tau$ in the form
\begin{eqnarray}
\tmH_\tau&=& (2\pi\alpha')
 p_{\ttheta} (\bV_\mu G^{\mu\nu}\bV_\nu +
\bX G_{\theta\theta})p_{\ttheta}+\nonumber \\
&+&(2\pi\alpha')\hPi_\mu G^{\mu\nu}\hPi_\nu +\hPi_\mu
(-2\partial_\sigma \ttheta G^{\theta\mu}-2\bV_\nu
\partial_\sigma x^\nu G^{\theta\mu})+ 4\pi\alpha'\hPi_\mu
 G^{\mu\nu}\bV_\nu p_{\ttheta}+\nonumber\\
&+&p_{\ttheta}(-2\partial_\sigma \ttheta G^{\theta\mu}\bV_\mu
-2\bV_\nu
\partial_\sigma x^\nu G^{\theta\mu}\bV_\mu
+2\bX G_{\mu \theta}\partial_\sigma
x^\mu)+\nonumber \\
&+&\frac{1}{2\pi\alpha'}\left(G^{\theta\theta}(\partial_\sigma
\ttheta)^2 -2\bV_\mu
\partial_\sigma x^\mu G^{\theta\theta}\partial_\sigma \ttheta+
\partial_\sigma x^\mu (\bV_\mu
G^{\theta\theta}\bV_\nu)\partial_\sigma x^\nu  +
  \bX G_{\mu\nu}\partial_\sigma
x^\mu\partial_\sigma x^\nu\right) \nonumber \\
\end{eqnarray}
and hence we obtain
\begin{eqnarray}
\partial_\tau \ttheta &=&\pb{\ttheta,H}=4\pi\alpha'
\lambda_\tau (\bV_\mu G^{\mu\nu}\bV_\nu+\bX) G_{\theta\theta}
p_{\ttheta}+\nonumber \\
&+&4\pi\alpha' \hPi_\mu G^{\mu\nu}\bV_\nu -2 (\partial_\sigma
\ttheta G^{\theta\mu}\bV_\nu+ \bV_\sigma\partial_\sigma x^\sigma
G^{\theta\mu}\bV_\mu-
\bX G_{\mu\theta}\partial_\sigma x^\mu)+\lambda_\sigma \partial_\sigma \theta \ , \nonumber \\
\partial_\tau x^\mu&=&\pb{x^\mu,H}=2\lambda_\tau((2\pi\alpha') G^{\mu\nu}\hPi_\mu
-(\partial_\sigma \ttheta +\bV_\nu\partial_\sigma
x^\nu)G^{\theta\mu}+(2\pi\alpha')G^{\mu\nu}\bV_\nu
p_{\ttheta})+\lambda_\sigma
\partial_\sigma x^\mu \  \ .
\nonumber \\
\end{eqnarray}
Now from  the last equation we get
\begin{equation}
\hPi_\mu=
\frac{1}{2\lambda_\tau}h_{\mu\nu}\bX^\nu-\frac{G_{\mu\theta}}{G_{\theta\theta}}
(\partial_\sigma \ttheta+\bV_\nu\partial_\sigma x^\nu) -\bV_\mu
p_{\ttheta} \ ,
%\frac{1}{2\lambda_\tau}\bX^\mu+ (\partial_\sigma
%\ttheta+\bV_\nu\partial_\sigma x^\nu)g^{\theta\mu}
%-(2\pi\alpha')g^{\mu\nu}\bV_\nu
%p_{\ttheta}=(2\pi\alpha')g^{\mu\nu}\hPi_\nu \ ,
\end{equation}
where
%%We multiply given equation with $\bV_\nu$ so that we obtain
%%\begin{equation}
%%\frac{1}{2\lambda_\tau}\bX^\nu \bV_\nu+ (\partial_\sigma \ttheta+
%%\bV_\nu \partial_\sigma x^\nu)g^{\theta\mu}\bV_\mu=
%%g^{\mu\nu}(2\pi\alpha')\hPi_\nu \bV_\mu+(2\pi\alpha')\bV_\mu
%%g^{\mu\nu}\bV_\nu p_{\ttheta}
%\end{equation}
%and insert to the equation for $\Theta$ so that we find
\begin{equation}
p_{\ttheta}=\frac{1}{4\pi\alpha'\lambda_\tau G_{\theta\theta}\bX}
(\Theta+\bX^\mu \bV_\mu+2\lambda_\tau \bX
G_{\mu\theta}\partial_\sigma x^\mu) \ ,
\end{equation}
and where $\Theta,\bX^\mu$ are defined in (\ref{defXmu}). If we then
proceed in the same way as in case of the fundamental string we
derive final form of the dual Lagrangian density
%Now using the equations
%\begin{equation}
%h_{\mu\nu}=g_{\mu\nu}-\frac{g_{\mu\theta}g_{\nu\theta}}{g_{\theta\theta}}
%\ , h_{\mu\nu}g^{\nu\rho}=\delta_\mu^\rho \ ,
%g^{\theta\mu}h_{\mu\nu}= -\frac{g_{\theta\nu}}{g_{\theta\theta}} \ .
%\end{equation}
%Now the Lagrangian density has the form
\begin{eqnarray}\label{defmL}
\tilde{\mL}&=&\partial_\tau x^\mu p_\mu+\partial_\tau \ttheta
p_{\ttheta}-H=
\nonumber \\
&=&T_{D1}\frac{1}{4\lambda_\tau}\left(\tg_{\tau\tau}-2\lambda_\sigma
\tg_{\tau\sigma}+ \lambda^2_\sigma
\tg_{\sigma\sigma}\right)-T_{D1}\lambda_\tau \bX
\tg_{\sigma\sigma}+\nonumber \\
&+&T_{D1} \left(\pi^\sigma
B_{\mu\nu}+C_{\mu\nu}^{(2)}-\frac{G_{\mu\theta}}{G_{\theta\theta}}(\pi^\sigma
B_{\mu\theta}+C^{(2)}_{\nu\theta})+
\frac{G_{\nu\theta}}{G_{\theta\theta}}(\pi^\sigma
B_{\mu\theta}+C^{(2)}_{\mu\theta})\partial_\tau
x^\mu \partial_\sigma x^\nu+\right.\nonumber \\
&+&\left.\frac{G_{\mu\theta}}{G_{\theta\theta}}\partial_\tau
\ttheta\partial_\sigma x^\mu -\frac{G_{\mu\theta}}{
G_{\theta\theta}}\partial_\tau x^\mu \partial_\sigma \ttheta\right)
\ ,
\nonumber \\
\end{eqnarray}
where
%using
%\begin{equation}
%\hPi_\mu=
%\frac{1}{2\lambda_\tau}h_{\mu\nu}\bX^\nu-\frac{g_{\mu\theta}}{g_{\theta\theta}}
%(\partial_\sigma \ttheta+\bV_\nu\partial_\sigma x^\nu) -\bV_\mu
%p_{\ttheta} \ .
%\end{equation}
%and where
\begin{eqnarray}
\tg_{\tau\tau}&=&\partial_\tau x^\mu \left(h_{\mu\nu}+\frac{\bV_\mu
\bV_\nu} {G_{\theta\theta}\bX}\right)\partial_\tau
x^\nu+\frac{1}{G_{\theta\theta}\bX}(\partial_\tau
\ttheta)^2+\frac{2}{G_{\theta\theta}\bX}
\partial_\tau \ttheta \partial_\tau x^\mu \bV_\mu  \ , \nonumber \\
\tg_{\tau\sigma}&=&\partial_\tau x^\mu
\left(h_{\mu\nu}+\frac{\bV_\mu \bV_\nu}
{G_{\theta\theta}\bX}\right)\partial_\sigma x^\nu+
\frac{1}{G_{\theta\theta}\bX}\partial_\tau \ttheta
\partial_\sigma \ttheta+\frac{2}{G_{\theta\theta}\bX}
\partial_\tau \ttheta \partial_\sigma x^\mu \bV_\mu
+\frac{2}{G_{\theta\theta}\bX}
\partial_\sigma \ttheta \partial_\tau x^\mu \bV_\mu
  \ , \nonumber \\
\tg_{\sigma\sigma}&=&\partial_\sigma x^\mu
\left(h_{\mu\nu}+\frac{\bV_\mu \bV_\nu}
{G_{\theta\theta}\bX}\right)\partial_\sigma x^\nu
+\frac{2}{G_{\theta\theta}\bX}
\partial_\tau \ttheta \partial_\sigma x^\mu \bV_\mu
+\frac{1}{G_{\theta\theta}\bX}(\partial_\sigma \ttheta)^2
  \ . \nonumber \\
\end{eqnarray}
Finally we eliminate $\lambda_\tau,\lambda_\sigma $ using
corresponding equations of motion
\begin{eqnarray}
%\lambda_\sigma:  \ g_{\tau\sigma}+\lambda_\sigma g_{\sigma\sigma}=0
%\Rightarrow
\lambda_\sigma=-\frac{\tg_{\tau\sigma}}{\tg_{\sigma\sigma}} \ ,
\quad
%\ , \nonumber \\
%-\frac{1}{4\lambda_\tau^2}\frac{g_{\tau\tau}g_{\sigma\sigma}-g_{\tau\sigma}^2}
%{g_{\sigma\sigma}}-g_{\sigma\sigma}\bX=0 \Rightarrow
 \lambda_\tau=
\frac{1}{2\sqrt{\bX} \tg_{\sigma\sigma}}
\sqrt{\tg_{\tau\sigma}^2-\tg_{\tau\tau}\tg_{\sigma\sigma}} \ .
\nonumber
\\
\end{eqnarray}
Inserting back to the Lagrangian (\ref{defmL}) we obtain final form
of the dual Lagrangian density
\begin{eqnarray}\label{LD1dual}
\mL&=&-T_{D1}\sqrt{e^{-2\Phi}+(\pi-C^{(0)})^2} \sqrt{-\det
\tilde{g}_{\alpha\beta}}+\nonumber \\
&+&T_{D1}\left((\pi \tilde{B}_{\mu\nu}+\tilde{C}^{(2)}_{\mu\nu})
\partial_\tau x^\mu \partial_\sigma x^\nu+
\frac{G_{\mu\theta}}{G_{\theta\theta}}\partial_\tau \ttheta
\partial_\sigma  x^\mu-\frac{G_{\mu\theta}}{G_{\theta\theta}}\partial_\tau
x^\mu\partial_\sigma \ttheta\right) \ ,
\nonumber \\
\end{eqnarray}
where
\begin{eqnarray}
\tG_{\mu\nu}&=&
G_{\mu\nu}-\frac{G_{\mu\theta}G_{\nu\theta}}{G_{\theta\theta}}
+\frac{(\pi B_{\mu\theta}+C_{\mu\theta}^{(2)})(\pi B_{\nu\theta}+
C^{(2)}_{\nu\theta})}{G_{\theta\theta}(e^{-2\Phi}+(\pi-C^{(0)})^2)} \ , \nonumber \\
\tilde{G}_{\ttheta\ttheta}&=&\frac{1}{G_{\theta\theta}
(e^{-2\Phi}+(\pi-C^{(0)})^2)}  \ , \nonumber \\
\tilde{G}_{\ttheta\mu}&=&\tG_{\mu\ttheta}=\frac{1}{G_{\theta\theta}(e^{-2\Phi}+(\pi-C^{(0)})^2)}
(\pi B_{\mu\theta}+C^{(2)}_{\mu\theta}) \ , \nonumber \\
 \tilde{B}_{\mu\nu}&=&
B_{\mu\nu}-\frac{G_{\mu\theta}B_{\nu\theta}}{G_{\theta\theta}}
+\frac{G_{\nu\theta}B_{\mu\theta}}{G_{\theta\theta}} \ . \nonumber
\\
\tC^{(2)}_{\mu\nu}&=&C^{(2)}_{\mu\nu}-\frac{G_{\mu\theta}}{G_{\theta\theta}}
C_{\nu\theta}^{(2)}+\frac{G_{\nu\theta}}{G_{\theta\theta}}C^{(2)}_{\mu\theta}
\ , \quad
%\tC^{(2)}_{\ttheta\mu}=-\tC^{(2)}_{\mu\ttheta}=\frac{G_{\mu\theta}}{G_{\theta\theta}}
%\ ,
 \nonumber \\
\end{eqnarray}
We would like to give physical interpretation of given
transformation rules.  To begin with note that  Type IIB theory is
invariant under $SL(2,Z)$ symmetry
\begin{eqnarray}
\hat{G}_{MN}&=&e^{\frac{1}{2}(\hat{\Phi}-\Phi)}G_{MN} \ , \quad
\hat{\tau}=
\frac{p\tau+q}{r\tau+s} \ , \nonumber \\
\hat{B}_{MN}&=&sB_{MN}-r C_{MN}^{(2)} \ , \quad \hat{C}^{(2)}_{MN}
=pC_{MN}-q B_{MN} \
, \nonumber \\
\end{eqnarray}
where $\tau=C^{(0)}+ie^{-\Phi}$ and where $ps-qr=1$.
 Let us presume
that $\pi=-|\pi|$ and then choose
 following values of the parameters $p,q,r,s$
\cite{Tseytlin:1996it} :
\begin{equation}
p=0 \ , q=-1 \ , r=1 \ , s=|\pi|
\end{equation}
so that we explicitly obtain
\begin{eqnarray}\label{rulesS}
\hat{C}^{(0)}&=&-\frac{C^{(0)}+|\pi|}{(C^{(0)}+|\pi|)^2+e^{-2\Phi}}
\ , \quad
e^{-\hat{\Phi}}=\frac{e^{-\Phi}}{(C^{(0)}+|\pi|)^2+e^{-2\Phi}} \ ,
\nonumber \\
\hat{G}_{MN}&=&\sqrt{(C^{(0)}+|\pi|)^2+e^{-2\Phi}}G_{MN} \ , \nonumber \\
\hat{B}_{MN}&=&|\pi| B_{MN}-C^{(2)}_{MN} \ , \quad \hat{C}^{(2)}_{MN}=B_{MN} \ .  \nonumber \\
\end{eqnarray}
We see that when we combine the square root
$\sqrt{(C^{(0)}+|\pi|)^2+e^{-2\Phi}}$ with $\sqrt{-\det \tg}$ and
use (\ref{rulesS}) we obtain that the Lagrangian density
(\ref{LD1dual}) has the form
\begin{eqnarray}
\mL=-\frac{1}{2\pi\alpha'}
\sqrt{-\det\bar{g}}+\frac{1}{2\pi\alpha'}\left( \bar{B}_{\mu\nu}
\partial_\tau x^\mu \partial_\sigma x^\nu+\bar{B}_{\ttheta\mu}\partial_\tau \ttheta\partial_\sigma
x^\mu+\bar{B}_{\mu\ttheta}\partial_\tau x^\mu
\partial_\sigma\ttheta\right) \ ,
\end{eqnarray}
where
\begin{equation}
 \bar{g}_{\alpha\beta}=\bar{G}_{\mu\nu}\partial_\alpha
x^\mu\partial_\beta x^\nu+\bar{G}_{\mu\ttheta}\partial_\alpha
x^\mu\partial_\beta \ttheta+\bar{G}_{\ttheta\nu}\partial_\alpha
\ttheta \partial_\beta x^\nu
+\bar{G}_{\ttheta\ttheta}\partial_\alpha \ttheta
\partial_\beta\ttheta \ ,
\end{equation}
and where the background fields $\bar{G}_{MN},\bar{B}_{MN}$ have
explicit form
\begin{eqnarray}
\bar{G}_{\mu\nu}&=&
\hat{G}_{\mu\nu}-\frac{\hat{G}_{\mu\theta}\hat{G}_{\nu\theta}}{\hat{G}_{\theta\theta}}
+\frac{\hat{B}_{\mu\theta}\hat{B}_{\nu\theta}}{\hat{G}_{\theta\theta}} \ , \nonumber \\
\bar{G}_{\ttheta\ttheta}&=&\frac{1}{\hat{G}_{\theta\theta}}  \ ,
\quad
\bar{G}_{\ttheta\mu}=\bar{G}_{\mu\ttheta}=\frac{1}{\hat{G}_{\theta\theta}}
\hat{B}_{\mu\theta} \ , \nonumber \\
 \bar{B}_{\mu\nu}&=&
\hat{B}_{\mu\nu}-\frac{\hat{G}_{\mu\theta}\hat{B}_{\nu\theta}}{\hat{G}_{\theta\theta}}
+\frac{\hat{G}_{\nu\theta}\hat{B}_{\mu\theta}}{\hat{G}_{\theta\theta}}
\ , \nonumber
\\
\bar{B}_{\ttheta\mu}&=&-\hat{B}_{\mu\ttheta}=\frac{\hat{G}_{\mu\theta}}{\hat{G}_{\theta\theta}}
\ .
 \nonumber \\
\end{eqnarray}
Now the physical interpretation of the canonical transformation
(\ref{canD1}) on the world-volume of D1-brane is clear. It
corresponds to the T-duality rules for the fundamental string moving
in the S-dual background (\ref{rulesS}).

%\begin{eqnarray}
%\tilde{\mL}&=&-T_{D1}\sqrt{e^{-2\Phi}+(\pi^\sigma-C^{(0)})^2}
%\sqrt{-\det \tg}+T_{D1}(\pi^\sigma
%\tilde{B}_{\mu\nu}+\tilde{C}^{(2)}_{\mu\nu})
%\partial_\tau x^\mu \partial_\sigma x^\nu+\nonumber \\
%&+& (\pi^\sigma\tB_{\mu\ttheta}+C^{(2)}_{\mu \ttheta})\partial_\tau
%x^\mu
%\partial_\sigma \ttheta+
%(\pi^\sigma\tB_{\ttheta\mu}+
% C^{(2)}_{\ttheta\mu})\partial_\tau
%\ttheta\partial_\sigma x^\mu \ ,
%\nonumber \\
%\end{eqnarray}
%where
%\begin{eqnarray}
%\tG_{\mu\nu}&=&G_{\mu\nu}-\frac{G_{\mu\theta}G_{\nu\theta}}{G_{\theta\theta}}
%+\frac{(\pi^\sigma B_{\mu\theta}+C_{\mu\theta}^{(2)})(\pi^\sigma
%B_{\nu\theta}+
%C^{(2)}_{\nu\theta})}{(e^{-2\Phi}+(\pi^\sigma-C^{(0)})^2)} \ , \nonumber \\
%\tB_{\mu\nu}&=&B_{\mu\nu}-\frac{G_{\mu\theta}B_{\nu\theta}}{G_{\theta\theta}}
%+\frac{G_{\nu\theta}B_{\mu\theta}}{G_{\theta\theta}} \ , \nonumber
%\\
%\tC^{(2)}_{\mu\nu}&=&
%C^{(2)}_{\mu\nu}-\frac{G_{\mu\theta}}{G_{\theta\theta}}
%C_{\nu\theta}^{(2)}+\frac{G_{\nu\theta}}{G_{\theta\theta}}C^{(2)}_{\mu\theta}
%\ , \nonumber \\
%\end{eqnarray}
%We see that canonical transformations performed on the world-volume
%of D1-brane gives the form of the Lagrangian where the
%transformation rules for the background fields do not coincide with
%the standard T-duality rules. This fact could be expected since
%D-branes behave under T-duality in more complicated ways than
%fundamental string.

\section{Canonical Transformations and Double Wick Rotation on D1-brane}\label{fifth}
In this section we perform the same analysis as in section
(\ref{third}) in order to find the relation between sequence of
canonical transformations and Wick rotation with the double Wick
rotation on the world-volume of gauge fixed D1-brane action. We
again consider the metric and NS-NS two form field in the form
(\ref{lightconemetric}) and we introduce the light cone coordinates
as in (\ref{defxplus}) so that the metric components are given in
(\ref{Gplusplus}) and (\ref{Gplusplusin}).
Further, using the relation between light-cone coordinates and the
original ones we obtain following components of Ramond-Ramond two
form
% In case of two
%form we have to be more careful and we have
%\begin{eqnarray}
%C=\frac{1}{2}C_{MN}dx^M\wedge dx^N= C_{t\varphi}dt\wedge d\varphi+
%C_{t\mu}dt\wedge dx^\mu+C_{\varphi \mu}d\varphi dx^\mu= \nonumber \\
%=C_{t\varphi}(dx^+-adx^-)\wedge (dx^++(1-a)dx^-)+\nonumber \\
%+(C^{(2)}_{t\mu}+C^{(2)}_{\varphi\mu})dx^+\wedge dx^\mu+
%(-C^{(2)}_{t\mu}+(1-a)C^{(2)}_{\varphi\mu})dx^-\wedge dx^\mu
%\nonumber
%\\
%%=\frac{1}{2}C_{t\varphi} dx^+\wedge
%%dx^--\frac{1}{2}C_{t\varphi}dx^-\wedge dx^+ \nonumber \\
%\end{eqnarray}
%and hence we have
\begin{eqnarray}
C^{(2)}_{+-}&=& -C^{(2)}_{-+}=C^{(2)}_{t\varphi} \ ,
%C^{(2)}_{-+}=-C^{(2)}_{t\varphi} \
 \nonumber \\
C^{(2)}_{+\mu}&=&-C_{\mu+}= C_{t\mu}^{(2)}+C_{\varphi\mu}^{(2)} \ ,
\nonumber \\
C_{-\mu}^{(2)}&=&-C^{(2)}_{\mu -}=
-C_{t\mu}^{(2)}+(1-a)C_{\varphi\mu}^{(2)} \ . \nonumber \\
\end{eqnarray}
In the light cone coordinates the  Hamiltonian and diffeomorphism
constraints have the form
\begin{eqnarray}
\mH_\tau&=&(2\pi\alpha')\Pi_+G^{++}\Pi_++2(2\pi\alpha')
\Pi_+G^{+-}\Pi_-+(2\pi\alpha')\Pi_-G^{--}\Pi_- +\nonumber \\
&+& \frac{1}{2\pi\alpha'}\bX \left(G_{++}(\partial_\sigma x^+)^2+
 2G_{+-}\partial_\sigma x^+\partial_\sigma
x^-+G_{--}(\partial_\sigma x^-)^2\right)+ \mH_x \ ,
 \nonumber \\
\mH_\sigma &=&p_+\partial_\sigma x^++p_-\partial_\sigma x^-+p_\mu
\partial_\sigma x^\mu \ ,  \nonumber \\
\end{eqnarray}
where
\begin{eqnarray}
 \mH_x &\equiv & (2\pi\alpha') \Pi_\mu G^{\mu\nu}\Pi_\nu
+ \frac{1}{2\pi\alpha'}\bX
\partial_\sigma x^\mu G_{\mu\nu}
\partial_\sigma x^\nu \ ,
\nonumber \\
\end{eqnarray}
and where $\bX=\left(e^{-2\Phi}+\left(C^{(0)}\right)^2\right)$. Note
that for  simplicity we presume that there is no electric flux so
that $\pi^\sigma=0$. Now we are ready to study the relation between
sequence of canonical transformation, target space Wick rotation and
world-sheet  double Wick rotation in the same way as in section
(\ref{third}). Let us presume that the background does not depend on
$x^-$ and perform the canonical transformation from $x^-$ to $\psi$
so we obtain the relation
%Let us presume that
%this generating function has the form
%\begin{equation}
%G(\theta,\psi)=\frac{1}{4\pi\alpha'}\int d\sigma (
%\partial_\sigma x^- \psi-x^- \partial_\sigma \psi)
%\end{equation}
%Let us denote the momentum conjugate to $\psi$ as $p_{\psi}$. Then
%we obtain
\begin{eqnarray}
p_{\psi}=-\frac{1}{2\pi\alpha'}\partial_\sigma x^-  \ , \quad p_{-}=
-\frac{1}{2\pi\alpha'}\partial_\sigma \psi \
\nonumber \\
\end{eqnarray}
and hence we find
% Now we obtain T-dual Hamiltonian when we replace
%$\partial_\sigma x^-$ with $-(2\pi\alpha') p_{\psi}$ and $p_-$ with
%$-\frac{1}{2\pi\alpha'}\partial_\sigma \psi$ and hence we find
\begin{eqnarray}
\Pi_+&=& p_+-\frac{1}{2\pi\alpha'} C^{(2)}_{+\mu}\partial_\sigma
x^\mu+C^{(2)}_{+-}p_{\psi} \ , \nonumber \\
\Pi_-&=&-\frac{1}{2\pi\alpha'}(\partial_\sigma \psi +
C_{-+}^{(2)}\partial_\sigma x^+)-\frac{1}{2\pi\alpha'}
C^{(2)}_{-\mu}\partial_\sigma x^\mu \ , \nonumber \\
\Pi_\mu&=&p_\mu-
\frac{1}{2\pi\alpha'}(C^{(2)}_{\mu\nu}\partial_\sigma x^\nu+
C^{(2)}_{\mu +}
\partial_\sigma x^+)+C_{\mu -}^{(2)}p_{\psi} \ .
\nonumber \\
\end{eqnarray}
%and hence
%\begin{eqnarray}
%\mH_\tau&=&(2\pi\alpha')\Pi_+G^{++}\Pi_++2(2\pi\alpha')
%\Pi_+G^{+-}\Pi_-+(2\pi\alpha')\Pi_-G^{--}\Pi_- +\nonumber \\
%&+& \frac{1}{2\pi\alpha'}
%\left(e^{-2\Phi}+\left(\pi^\sigma-C^{(0)}\right)^2\right)
%\left(G_{++}(\partial_\sigma x^+)^2+\right.\nonumber \\
%&-& \left. 2(2\pi\alpha')G_{+-}\partial_\sigma x^+
%p_{\psi}+(2\pi\alpha')^2G_{--}(p_\psi)^2\right)+ \mH_x \ ,
% \nonumber \\
%\mH_\sigma &=&p_+\partial_\sigma x^++p_\psi\partial_\sigma
%\psi+p_\mu
%\partial_\sigma x^\mu \ ,  \nonumber \\
%\end{eqnarray}
As the next step we perform analytic continuation
\begin{equation}
(x^+,\psi)\rightarrow (i\tpsi, -i\tx^+)
 \ , \quad (p_+,p_\psi)\rightarrow (-i\tp_\psi, i\tp_+)
 \end{equation}
that implies
\begin{eqnarray}
\Pi_+ &\rightarrow &
-i(\tp_\psi-T_{D1}C_{+-}^{(2)}\tp_+)-\frac{1}{2\pi\alpha'}C^{(2)}_{+\mu}
\partial_\sigma x^\mu
\ , \nonumber
\\
\Pi_-&\rightarrow & i(\partial_\sigma
\tx^+-T_{D1}C_{-+}^{(2)}\partial_\sigma
\tpsi)-\frac{1}{2\pi\alpha'}C^{(2)}_{-\mu}\partial_\sigma x^\mu \ , \nonumber \\
\Pi_\mu &\rightarrow& p_\mu -T_{D1}C_{\mu\nu}^{(2)}\partial_\sigma
x^\nu -iT_{D1}C_{\mu+}^{(2)}\partial_\sigma \tpsi+i
T_{D1}C^{(2)}_{\mu -}\tp_+ \ , \nonumber \\
\end{eqnarray}
 so that the Hamiltonian and spatial diffeomorphism constraints
 take the form
\begin{eqnarray}
& &\tmH_\tau=-(2\pi\alpha')
(\tp_\psi-T_{D1}C_{+-}^{(2)}\tp_+)^2 G^{++} +\nonumber \\
&+& 2(\tp_\psi-T_{D1}C_{+-}^{(2)}\tp_+)G^{+-} (\partial_\sigma
\tx^+-T_{D1}C_{-+}^{(2)}\partial_\sigma \tpsi)
-\frac{1}{2\pi\alpha'}(\partial_\sigma
\tx^+-T_{D1}C_{-+}^{(2)}\partial_\sigma \tpsi)^2 G^{--}
+\nonumber \\
&+& \frac{1}{2\pi\alpha'}\bX(-G_{++}(\partial_\sigma \tpsi)^2+
4\pi\alpha'G_{+-}\partial_\sigma \tpsi \tp_+-(2\pi\alpha')^2
G_{--}(\tp_+)^2+ G_{\mu\nu}
(\partial_\sigma x^\mu \partial_\sigma x^\nu)+\nonumber \\
&+&(p_\mu -\frac{1}{2\pi\alpha'}(C_{\mu\sigma}^{(2)}\partial_\sigma
x^\sigma -iC^{(2)}_{\mu+}\partial_\sigma \tpsi +i(2\pi\alpha')C_{\mu
-}^{(2)} \tp_+))G^{\mu\nu}\times \nonumber \\
&\times & (p_\nu -\frac{1}{2\pi\alpha'}(
C_{\nu\rho}^{(2)}\partial_\sigma x^\rho
-iC^{(2)}_{\nu+}\partial_\sigma \tpsi +i(2\pi\alpha')C_{\nu
-}^{(2)} \tp_+)) \ , \nonumber \\
& & \tmH_\sigma=\tpsi \partial_\sigma \tp_{\psi}+
\tx^+\partial_\sigma
\tp_++p_\mu \partial_\sigma x^\mu  \ .  \nonumber \\
\end{eqnarray}
As the third step  we perform canonical transformation along $\tpsi$
direction. We denote dual coordinate as $\tphi$ so that we have
\begin{equation}
\partial_\sigma \tpsi= -(2\pi\alpha')p_{\tphi} \ , \quad
\tp_\psi=-\frac{1}{2\pi\alpha'}\partial_\sigma \tphi \ .
\end{equation}
Inserting these relations to $\tmH_\tau,\tmH_\sigma$ given above  we
derive the final form of the Hamiltonian and diffeomorphism
constraints
\begin{eqnarray}
& & \tmH_\tau=-(2\pi\alpha')
(\frac{1}{2\pi\alpha'}\partial_\sigma\tphi+
C_{+-}^{(2)}\tp_+)^2 G^{++} -\nonumber \\
&-& \frac{2}{2\pi\alpha'}
 (\partial_\sigma \tphi+(2
\pi\alpha')C_{+-}^{(2)}\tp_+)G^{+-} (\partial_\sigma
\tx^++(2\pi\alpha')C_{-+}^{(2)} p_{\tphi}) -\frac{1}{2\pi\alpha'}
(\partial_\sigma \tx^++(2\pi\alpha')C_{-+}^{(2)} p_{\tphi})^2 G^{--}
+\nonumber \\
&+& (2\pi\alpha')\bX(-G_{++}(p_{\tphi})^2- 2G_{+-}p_{\tphi}
\tp_+-G_{--}(\tp_+)^2+ \frac{1}{(2\pi\alpha')^2}G_{\mu\nu}
(\partial_\sigma x^\mu \partial_\sigma x^\nu))+\nonumber \\
&+&(p_\mu -\frac{1}{2\pi\alpha'} C_{\mu\sigma}^{(2)}\partial_\sigma
x^\sigma +iC^{(2)}_{\mu+}p_{\tphi} +iC_{\mu -}^{(2)}
\tp_+)G^{\mu\nu}
 (p_\nu
-\frac{1}{2\pi\alpha'}C_{\nu\rho}^{(2)}\partial_\sigma x^\rho
+iC^{(2)}_{\nu+}p_{\tphi} +iC_{\nu
-}^{(2)} \tp_+) \ , \nonumber \\
& & \tmH_\sigma =\partial_\sigma \tphi  p_{\tphi}+
\partial_\sigma \tx^+
\tp_++p_\mu \partial_\sigma x^\mu  \ .  \nonumber \\
\end{eqnarray}
Finally we fix the gauge by imposing the constraints
\begin{equation}
p_{\tphi}=\frac{1}{2\pi\alpha'}  \ , \quad  x^+=\tau \ .
\end{equation}
Then we obtain that the Hamiltonian for the physical degrees of
freedom should be identified as
\begin{equation}
\mH_{fix}=-p_+ \ .
\end{equation}
From $\tmH_\sigma$ we obtain
$\partial_\sigma\tphi=-(2\pi\alpha')\partial_\sigma x^\mu p_\mu$.
Further we see that in order to have real Hamiltonian we have to
demand that $C^{(2)}_{\mu +}=C^{(2)}_{\mu -}=0$. Then the strong
constraint $\tmH_\tau=0$ is equal to
\begin{eqnarray}
\mH_\tau&=&-(2\pi\alpha') (p_\mu \partial_\sigma x^\mu-
C_{+-}^{(2)}\tp_+)^2 G^{++} +\nonumber \\
 &+&2
 (\partial_\sigma x^\mu p_\mu-C_{+-}^{(2)}\tp_+)G^{+-}C^{(2)}_{-+}
 -\frac{1}{2\pi\alpha'}  G^{--}(C^{(2)})^2_{-+}
+\nonumber \\
&+& (2\pi\alpha')\bX(-G_{++}\frac{1}{(2\pi\alpha')^2}-
\frac{2}{2\pi\alpha'}G_{+-} \tp_+-G_{--}(\tp_+)^2)+ \mH_x \ .
\nonumber \\
\end{eqnarray}
This equation can be solved for $\tp_+$ and we obtain
%\begin{eqnarray}
%-(2\pi\alpha')\tp_+^2(G_{--}\bX+(C_{+-}^{(2)})^2G^{++})-2
%\tp_+(G_{+-}\bX-(C_{+-}^{(2)})^2G^{+-}-2\pi\alpha'p_\mu
%\partial_\sigma x^\mu
%C_{+-}^{(2)}G^{++})- \nonumber \\
%-(2\pi\alpha')(p_\mu\partial_\sigma x^\mu)^2 G^{++}+
%2\partial_\sigma x^\mu p_\mu
%G^{+-}C_{-+}^{(2)}-\frac{1}{2\pi\alpha'}G^{--}C_{-+}^2
%-\frac{1}{2\pi\alpha'}\bX G_{++}+\mH_x=0 \nonumber \\
%\end{eqnarray}
%and hence we obtain
\begin{eqnarray}\label{p+TWT}
\tp_+=\frac{\tG^{+-}}{2\pi\alpha'\tG^{++}}-\frac{2}{2\pi\alpha'\tG^{++}}\sqrt{\bK}
+ p_\mu\partial_\sigma x^\mu
\tC^{(2)}_{+-} \ , \nonumber \\
\end{eqnarray}
where
\begin{eqnarray}
\bK=
%4 [(G_{+-}\bX-C_{+-}^2 G^{+-})^2+(2\pi\alpha')^2 (p_\mu
%\partial_\sigma x^\mu)^2 C_{+-}^2 G^{++}G^{++}]-\nonumber \\
%-4(2\pi\alpha')(-(G_{--}\bX+C_{+-}^2 G^{++})) (-(2\pi\alpha') (p_\mu
%\partial_\sigma x^\mu)^2G^{++}-\frac{1}{2\pi\alpha'}G^{--}C_{+-}^2
%-\frac{1}{2\pi\alpha'}\bX G_{++}+\mH_x)=
%\nonumber \\
(\tG^{+-})^2-\tG^{++}\tG^{--}-(2\pi\alpha')^2\tG^{++}\tG_{--}\bX(p_\mu
\partial_\sigma x^\mu)^2
-(2\pi\alpha')\tG^{++}\mH_x \ , \nonumber \\
\end{eqnarray}and
where we identified the new background fields
\begin{eqnarray}\label{backfieldD1}
\tC_{+-}^{(2)}&=&-\frac{C_{+-}^{(2)}G^{++}}{G_{--}\bX+(C_{+-}^{(2)})^2
G^{++}} \ , \nonumber \\
\tG^{++}&=&-\bX G_{--}-(C_{+-}^{(2)})^2 G^{++} \ , \nonumber \\
\tG^{+-}&=&G_{+-}\bX-(C^{(2)}_{+-})^2G^{+-} \ , \nonumber \\
\tG^{--}&=&-G^{--}(C^{(2)}_{+-})^2-\bX G_{++} \ . \nonumber \\
\end{eqnarray}
As in section (\ref{third}) we compare this result with the uniform
gauge fixed D1-brane when we consider  canonical dual Hamiltonian
constraint
\begin{eqnarray}
\mH_\tau&=&(2\pi\alpha')\Pi_+G^{++}\Pi_++2(2\pi\alpha')
\Pi_+G^{+-}\Pi_-+(2\pi\alpha')\Pi_-G^{--}\Pi_- +\nonumber \\
&+& \frac{1}{2\pi\alpha'} \bX \left(G_{++}(\partial_\sigma x^+)^2-
 4\pi\alpha'G_{+-}\partial_\sigma x^+
p_{\psi}+(2\pi\alpha')^2G_{--}(p_\psi)^2\right)+ \mH_x \ ,
 \nonumber \\
\mH_\sigma &=&p_+\partial_\sigma x^++p_\psi\partial_\sigma
\psi+p_\mu
\partial_\sigma x^\mu \ ,  \nonumber \\
\end{eqnarray}
where
\begin{eqnarray}
\Pi_+=p_++C^{(2)}_{+-}p_{\psi} \ , \quad
\Pi_-=-\frac{1}{2\pi\alpha'}(\partial_\sigma \psi +
C_{-+}^{(2)}\partial_\sigma x^+)\ , \quad \Pi_\mu=p_\mu-
\frac{1}{2\pi\alpha'}C^{(2)}_{\mu\nu}\partial_\sigma x^\nu\ .
\nonumber \\
\end{eqnarray}
Then we perform the gauge fixing
\begin{equation}
x^+=\tau \ , \psi=\sigma \
\end{equation}
and  we obtain
\begin{equation}
\Pi_-=-\frac{1}{2\pi\alpha'} \ , \quad p_{\psi}= -p_\mu
\partial_\sigma x^\mu
\end{equation}
and hence the Hamiltonian constraint is equal to
\begin{eqnarray}
0&=&\mH_\tau=(2\pi\alpha')\Pi_+G^{++}\Pi_+-2
\Pi_+G^{+-}+\frac{1}{2\pi\alpha'}G^{--} +\nonumber \\
&+& (2\pi\alpha')\bX G_{--}(p_\mu\partial_\sigma x^\mu)^2+ \mH_x \
 \nonumber \\
\end{eqnarray}
that can be solved for $p_+$ as
\begin{eqnarray}\label{p+}
& &p_+=\frac{G^{+-}}{2\pi\alpha'G^{++}}-
+C_{+-}^{(2)}(p_\mu\partial_\sigma x^\mu)
\nonumber \\
&-&\frac{2}{2\pi\alpha'G^{++}} \sqrt{((G^{+-})^2-G^{++}G^{--})
-(2\pi\alpha')^2G^{++}G_ {--} \bX(p_\mu
\partial_\sigma x^\mu)^2- (2\pi\alpha')
G^{++}\mH_x} \ .  \nonumber \\
\end{eqnarray}
Now comparing (\ref{p+}) with (\ref{p+TWT}) we see that these two
gauge fixed Hamiltonians have the same form when we perform the
identification of the background fields as was given in
(\ref{backfieldD1}). It is important to stress that the
transformation rules (\ref{backfieldD1}) coincide with the rules
that were derived in \cite{Kluson:2015saa} when the double Wick
rotation was performed on the world-volume of uniform gauge fixed
D1-brane action. We also see that $\bX$ does not transform again
with agreement with \cite{Kluson:2015saa}. Finally using the
arguments given in section (\ref{fourth}) we can argue that
(\ref{backfieldD1}) are in agreement with \cite{Arutyunov:2014cra}
when $C_{+-}^{(2)}=0$. Then we can rewrite (\ref{backfieldD1}) into
the form
\begin{eqnarray}\label{backfieldD1red}
\frac{1}{\sqrt{(C^{(0)})^2+e^{-2\Phi}}}
\tG^{++}&=&- \sqrt{e^{-2\Phi}+(C^{(0)})^2} G_{--} \ , \nonumber \\
\frac{1}{\sqrt{(C^{(0)})^2+e^{-2\Phi}}}\tG^{+-}&=&G_{+-}
\sqrt{(C^{(0)})^2+e^{-2\Phi}} \ ,  \nonumber \\
\frac{1}{\sqrt{(C^{(0)})^2+e^{-2\Phi}}}
\tG^{--}&=&\sqrt{(C^{(0)})^2+e^{-2\Phi}} G_{++} \ . \nonumber \\
\end{eqnarray}
We immediately see that these metric components correspond to the
S-dual metric when we used the equivalence between D1-brane action
and the fundamental string action in S-dual background. Hence
(\ref{backfieldD1red}) precisely correspond to the rules derived in
\cite{Arutyunov:2014cra} when are applied for the string moving in
S-dual background. 

 In summary, we have
shown that there is a equivalence between sequence of canonical
transformations and target space Wick rotation on one side  and the
double Wick rotation on the gauge fixed D1-brane action on the
another side. On the other hand we also see that the components of
the two form field $C^{(2)}_{\mu\nu}$ that are transverse to the
directions where the canonical transformations were performed do not
transform while their change the sign in case of double Wick
rotation of the uniform gauge fixed D1-brane action. On the other
hand we can argue as in section (\ref{second}) that the
transformation $C^{2}\rightarrow -C^{(2)}$ maps one solutions of the
supergravity equations of motion to another one (together with
$B\rightarrow -B$) and hence we see that there is an equivalence
between double Wick rotation on the world-volume of uniform gauge
fixed D1-brane and sequence of transformations: "canonical
transformation-target space double Wick rotation-canonical
transformation-$C^{(2)}\rightarrow -C^{(2)},B\rightarrow -B$).

 \vskip .5in
\noindent {\bf Acknowledgement:}

 This work   was
supported by the Grant agency of the Czech republic under the grant
P201/12/G028. \vskip 5mm

\end{document}